\documentclass[1p]{elsarticle}
\pdfoutput=1
\usepackage{amsmath}
\usepackage{graphicx}
\usepackage{caption}
\expandafter\def\csname ver@subfig.sty\endcsname{}
\usepackage{hyperref}
\usepackage{tikz}
\usepackage{pgfplots}
\usepackage{mathtools}
\pgfplotsset{compat=1.14}
\usepackage{amsfonts}
\usepackage{amssymb}
\usepackage{wasysym}
\pgfplotsset{plot coordinates/math parser=false}
\newlength\figureheight
\newlength\figurewidth
\usetikzlibrary{plotmarks}
\usetikzlibrary{calc}

 \pgfplotsset{
        layers/my layer set/.define layer set={
            background,
            main,
            foreground
        }{
        },
        set layers=my layer set,
    }

\begin{document}

\title{Theory and particle tracking simulations of a resonant radiofrequency deflection cavity in TM$_{110}$ mode for ultrafast electron microscopy}
\address[tue]{Department of Applied Physics, Coherence and Quantum Technology Group, Eindhoven University of Technology, P.O.~Box 513, 5600 MB Eindhoven, the Netherlands}
\address[fei]{Thermo Fisher Scientific, Achtseweg Noord 5, 5651 GG Eindhoven, the Netherlands}
\address[icms]{Institute for Complex Molecular Systems, Eindhoven University of Technology, P.O.~Box 513, 5600 MB Eindhoven, the Netherlands}
\author[tue]{J.F.M.~van Rens\corref{cor}}
\cortext[cor]{Corresponding author}
\ead{j.f.m.v.rens@tue.nl}
\author[tue]{W.~Verhoeven}
\author[tue,icms]{J.G.H.~Franssen}
\author[tue]{A.C.~Lassise}
\author[tue]{X.F.D.~Stragier}
\author[fei]{E.R.~Kieft}
\author[tue]{P.H.A.~Mutsaers}
\author[tue,icms]{O.J.~Luiten}
\date{\today}

\begin{abstract}
We present a theoretical description of resonant radiofrequency (RF) deflecting cavities in TM$_{110}$ mode as dynamic optical elements for ultrafast electron microscopy. We first derive the optical transfer matrix of an ideal pillbox cavity and use a Courant-Snyder formalism to calculate the 6D phase space propagation of a Gaussian electron distribution through the cavity. We derive closed, analytic expressions for the increase in transverse emittance and energy spread of the electron distribution. We demonstrate that for the special case of a beam focused in the center of the cavity, the low emittance and low energy spread of a high quality beam can be maintained, which allows high-repetition rate, ultrafast electron microscopy with 100 fs temporal resolution combined with the atomic resolution of a high-end TEM. This is confirmed by charged particle tracking simulations using a realistic cavity geometry, including fringe fields at the cavity entrance and exit apertures.
\end{abstract}

\maketitle

\section{Introduction}

Since the introduction of the Ultrafast (Transmission) Electron Microscope (U(T)EM) by Ahmed Zewail \cite{Lobastov2005}, the dynamics of various sorts of material properties have been studied using ultrafast electron techniques such as imaging \cite{Zewail2010} \cite{Flannigan2012}, diffraction \cite{Sciaini2011} and electron energy-loss spectroscopy (EELS) \cite{Carbone2009} \cite{VanderVeen2015} with picosecond to femtosecond temporal resolution. The research described in these references is all based on a pump-probe scheme in which the probing electron pulses are created from a flat photo-cathode using a fs laser system. This causes two limitations: First, the average current of the UTEM is limited by the repetition rate of the fs laser system, although long relaxation times of dynamical processes, or slow thermal diffusion, can also limit the maximal repetition times that can be used. Second, the relatively large area of the flat-photocathode limits the peak brightness of the generated electron pulses, hence the maximally achievable spatial resolution. A significant improvement in the peak brightness of laser-triggered electron sources has been achieved by sideways laser illumination of a nano-tip. The reduced dimensions of the photo-field emitter have resulted in a working UTEM with 200 fs electron pulses with a peak brightness comparable to continuous Schottky sources \cite{Feist2017}. This technique has resulted in very impressive results \cite{Ehberger2015} \cite{Feist2015} \cite{Echternkamp2016}.

An alternative approach requiring no laser at all, involves the chopping of a continuous beam of a high-end TEM into ultrashort electron pulses using a fast blanker in combination with a slit. Apart from the lack of need for an amplified laser system to create electron pulses, further advantages are that no alterations are needed to the gun and the fact that it allows easy switching between continuous mode and pulsed mode. The principle of chopping an electron beam has been realized in Scanning Electron Microscopes (SEMs) many years ago \cite{Fujioka1983} \cite{Thong1991} in the form of electrostatic blanking capacitors \cite{Winkler1990} \cite{Fehr1990} and cavity resonators \cite{Oldfield2001} \cite{Hosokawa1978}. More recently, the use of a photo-conducting switch was proposed to create a laser-triggered, electrostatic beam blanker which can be used for pump-probe experiments in a SEM \cite{Weppelman2018}. In parallel, advanced RF-laser synchronization techniques \cite{Kiewiet2002} \cite{Gliserin2013} have reduced the timing jitter between electron pulses and laser pulses to levels below 100 fs \cite{Brussaard2013} and even 5 fs \cite{Walbran2015}, making RF cavity-based pulsed beams also suitable for pump-probe experiments.

RF cavities or resonators are specifically tailored metallic structures, in which electromagnetic energy can be stored in standing waves or modes. Because of resonant enhancement, RF cavities can be used to generate EM fields of high amplitudes with relatively low input power. Various types of RF cavities have been important elements of the standard toolbox for particle accelerators for many years. For example, a cavity in TM$_{010}$-mode supports an oscillating, electric field pointing along the beam axis, which is commonly used for the acceleration of relativistic charged particle pulses. Synchronized to a mode-locked laser system, a cavity in TM$_{010}$ mode can be used to for the compression of electron pulses in ultrafast electron diffraction experiments, resulting in pulses shorter than 100 fs \cite{VanOudheusden2010}. A cavity in TM$_{110}$ mode supports a magnetic field oscillating perpendicular to the beam axis, transversely deflecting the beam. This mode has been used to chop the continuous beam of a 30 kV SEM into ultrashort pulses \cite{Lassise2012} and record time-of-flight electron energy loss spectra \cite{Verhoeven2016}. Synchronized to a mode-locked laser, a cavity in TM$_{110}$ mode can be used for pulse length measurements \cite{VanOudheusden2010}, for example of non-relativistic, ultracold electron pulses extracted from laser cooled gases \cite{Franssen2017}. Note the same principles of pulse compression and metrology have also been applied with single-cycle THz fields instead of RF cavities \cite{Kealhofer2016}.

In 2012, Lassise \emph{et al.} showed that using a miniaturized RF cavity in TM$_{110}$ mode, it is possible to chop a 30 kV electron beam while fully maintaining the peak brightness \cite{Lassise2012PhDthesis}. Moreover he proposed to use this technique to chop the beam of a high-quality beam of a 200 kV TEM, also conserving the peak brightness of the Schottky field emission source \cite{Lassise2012PhDthesis}. Since then, such an RF cavity-based UTEM has been built at Eindhoven University of Technology (TU/e) and is currently operational \cite{Verhoeven2017}. Furthermore, alternative TEM beam chopping schemes involving multiple RF cavities are being investigated elsewhere \cite{Qiu}. For successful implementation of an RF cavity in a charged particle beam line, a thorough understanding of its effect on the beam dynamics is essential. If not used properly, the rapidly oscillating non-uniform and strong EM fields in RF cavities can have a detrimental effect on the beam quality. However, with proper settings of experimental parameters, such as the RF phase and the position of the beam crossover, the quality of the original beam can be fully maintained, essential for applications such as electron microscopy.

In this paper we present the theoretical background of a resonant RF cavity in TM$_{110}$ mode as a dynamic optical element for UTEM. In section \ref{sec:trajectories} we explain the principle of deflection and chopping by calculating the trajectories of a charged particle propagating through an ideal TM$_{110}$ pillbox cavity. From these trajectories, we derive the optical transfer matrix of the cavity in section \ref{sec:matrix}; and apply this in a Courant-Snyder formalism to calculate the 6D phase space propagation of a Gaussian electron distribution in section \ref{sec:CS}. In section \ref{sec:application} we apply our findings to study the special case of a focused beam inside a 200 kV TEM column. We derive closed, analytic expressions for the increase in transverse emittance and energy spread. We show that using proper experimental parameter settings, the growth in transverse emittance can be fully eliminated and also the increase in energy spread can be minimized. In section \ref{sec:simulations} we present charged particle tracking simulations using a realistic cavity geometry, including fringe fields at the cavity entrance and exit apertures, and a realistic electron beam. The simulations confirm our theoretical findings that an RF cavity in TM$_{110}$ mode can be used to chop an electron beam with negligible increase in both transverse emittance and energy spread. This property makes it a very interesting alternative for photo-emission based UTEM, especially in the light of the ever increasing brightness of continuous electron sources.

\section{Charged particle trajectories}\label{sec:trajectories}

\subsection{Beam brightness and emittance}\label{sec:merit}
\begin{figure}[htp]
\centering
\includegraphics[width=0.8\linewidth]{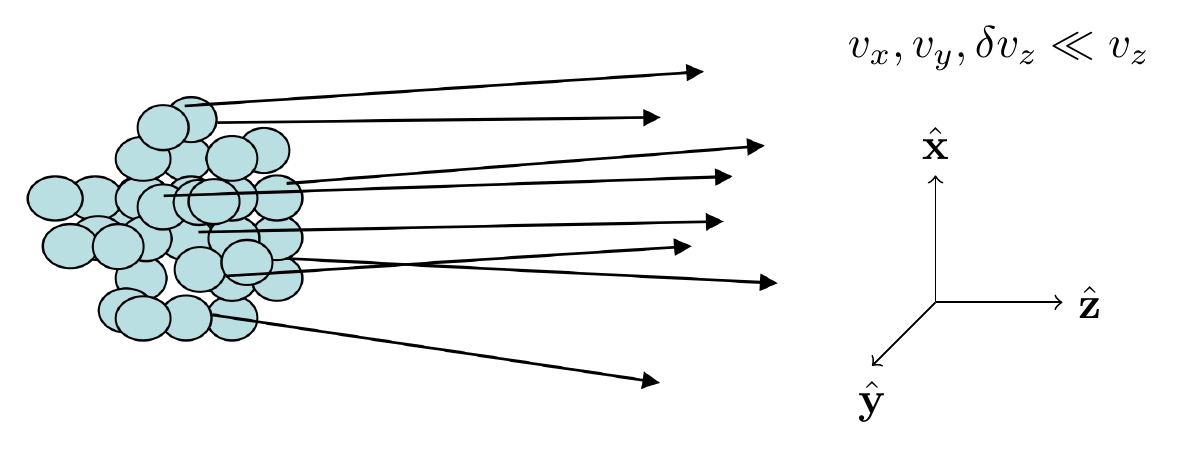}
\caption{A charged particle beam traveling in the $z$-direction. The velocity of any single particle is the sum of the mean velocity $v_z\hat{\mathbf{z}}$ and a small velocity $\delta \mathbf{v}=(v_x,v_y,\delta v_z)$ for which $v_x,v_y,\delta v_z \ll v_z$.}
\label{fig:beam}
\end{figure}

First, we define a beam as a distribution of charged particles (charge $q$ and mass $m$) with a collective motion along the $z$-axis with average speed $v_z$, for which each individual particle also has a small relative velocity $\delta \mathbf{v}=(v_x,v_y,\delta v_z)$, with
\begin{equation}\label{paraxial}
v_x,v_y,\delta v_z \ll v_z,
\end{equation}
see also figure \ref{fig:beam}.

We describe the motion of each particle in the distribution by the 6D trace space coordinates $x$, $x'$, $y$, $y'$, $z$ and $z'$, in which $x'\equiv\frac{v_x}{v_z}$ and $y'\equiv\frac{v_y}{v_z}$ are the angles of the particle trajectory with respect to the mean trajectory of the total distribution. Analogously, $z'\equiv \frac{\delta v_z}{v_z}$ represents the relative difference in longitudinal velocity of an individual particle with respect to the mean velocity of the beam. Note that we have used the paraxial approximation of equation \eqref{paraxial}. The 6D trace space volume occupied by this distribution is a measure for the quality of the beam. In accelerator physics, this is often expressed in terms of the root-mean-squared (rms) geometrical emittance in the $j=x,y,z$ direction \cite{Reiser}
\begin{equation}\label{eNx}
\epsilon_j\equiv\sqrt{\left< j^2\right> \left< j'^2\right> - \left< j j'\right>^2}.
\end{equation}
Here the brackets indicate averaging over the entire distribution. The emittance $\epsilon_j$ is proportional to the area of the projection of the 6D trace-space density on the ($j,j'$)-plane and is a measure for the focusability of the beam in the $j$-direction, given that the beam energy is fixed. The geometrical emittance $\epsilon_j$ is not a Lorentz-invariant quantity and therefore is not conserved during acceleration. To compare beams of different energy, the Lorentz-invariant rms normalized emittance in the $j=x,y,z$ direction is defined as
\begin{equation}\label{emittancex}
\epsilon_{n,j}\equiv\frac{\sqrt{\left< j^2\right> \left< p_j^2\right> - \left< j p_j \right>^2}}{mc}\approx \beta\gamma \epsilon_j,
\end{equation}
in which $p_j=\gamma m v_j$ is the momentum in the $j$-direction, $\gamma=1/\sqrt{1-\beta^2}$ is the relativistic Lorentz factor and $\beta=v/c$ is the normalized speed \cite{Reiser}. In a beam waist, there are no correlations between $x$ and $x'$, hence the normalized transverse emittance is simply given by
\begin{equation}\label{eNx}
\epsilon_{n,x}^{\textrm{waist}}\equiv\beta\gamma\sigma_x \sigma_{x'},
\end{equation}
in which $\sigma_x=\sqrt{\left< x^2\right>}$ and $\sigma_{x'}=\sqrt{\left< x'^2\right>}$ are the rms beam radius and rms semi-divergence angle in the beam waist. The quality of the beam in the $z$-direction is determined by the normalized longitudinal emittance $\epsilon_{n,z}$. For an charged particle bunch with no chirp $\left<z p_z\right>=0$, so the normalized longitudinal emittance $\epsilon_{n,z}$ can be written as
\begin{equation}
\epsilon_{n,z}=\frac{\sigma_z \sigma_{p_z}}{mc}\approx\frac{\sigma_t \sigma_U}{m c},
\end{equation}
in which $\sigma_t$ is the rms pulse duration and $\sigma_U$ is the rms energy spread \cite{Reiser}.

Electron microscopists often describe the quality of the beam in terms of the reduced brightness, which is locally defined as
\begin{equation}\label{brightness}
B_r\equiv \frac{1}{V^\ast}\frac{\mathrm{d}^2 I}{\mathrm{d}A\mathrm{d}\Omega}.
\end{equation}
The reduced brightness is proportional to the local current density $\frac{\mathrm{d} I}{\mathrm{d}A}=\frac{\mathrm{d}^2 I}{\mathrm{d}x\mathrm{d}y}$ per unit solid angle $\mathrm{d}\Omega=\mathrm{d}x'\mathrm{d}y'$. By dividing by the relativistically corrected beam potential $V^\ast \equiv (1/2+\gamma/2)V$ with $q V =(\gamma-1)m c^2$, this quantity is also Lorentz-invariant. To define a measure for the overall quality of a charged particle beam, the practical reduced brightness was introduced by Bronsgeest \emph{et al.} \cite{Bronsgeest2008}
\begin{equation}\label{Bpract}
B_{\mathrm{pract}}\equiv\frac{1}{V^\ast}\frac{I}{A_{\mathrm{pract}}\Omega}=\frac{1}{V^\ast}\frac{I}{\pi \left(\frac{d_{50}}{2}\right)^2 \pi \theta_x^2}.
\end{equation}
It defines the amount of current $I$ that can be focused into a waist with an area $A_{\mathrm{pract}}=\pi \left(\frac{d_{50}}{2}\right)^2$, in which $d_{50}$ is the full width spot diameter in which 50\% of the current is focused. Furthermore, it assumes a uniform angular distribution with semi divergence angle $\theta_x$. We can now express the practical brightness in terms of the rms normalized transverse emittance $\epsilon_{n,x}=\beta\gamma \sigma_x \sigma_{x'}$. By assuming a uniform, angular distribution so that $\sigma_{x'}=\theta_x/2$, and a Gaussian, spatial distribution in the beam waist so that $\sigma_x=d_{50}/2\sqrt{2\ln{2}}$, the practical reduced brightness can be expressed in terms of the rms, normalized transverse emittance as
\begin{equation}\label{Bvseps}
B_{\mathrm{pract}}=\frac{q}{m c^2}\frac{I}{4\ln{2} \cdot\pi^2 \epsilon_{n,x}^2}.
\end{equation}

\subsection{Framework and assumptions}\label{sec:framework}
Consider an ideal, cylindrical cavity in TM$_{110}$ mode of length $L_{}$, aligned along the $z$-axis of a cartesian coordinate system, and with the entrance aperture positioned at $(x,y,z)=(0,0,0)$. Close to the $z$-axis:
 \begin{equation}\label{kr}
 kx, ky\ll1,
 \end{equation}
 with $k$ the wavenumber of the RF field, the magnetic field $\mathbf{B}$ and the electric field $\mathbf{E}$ of a cavity in TM$_{110}$ mode can be approximated by
\begin{equation}\label{EMfields}
\left.
\begin{aligned}
\mathbf{B}&=B_{0}\cos(\phi_0+\omega t)\hat{\mathbf{y}} \\
\mathbf{E}&=-B_0 \omega x \sin(\phi_0+\omega t)\hat{\mathbf{z}}
\end{aligned}
\right\} \mathrm{for } ~0<z<L_{},
\end{equation}
in which $B_0$ is the magnetic field amplitude, $\omega=ck$ is the cavity resonance frequency and $\phi_0$ is the phase of the RF field at $t=0$ \cite{Lassise2012PhDthesis}. The word 'ideal' refers to the top-hat profile of the magnetic field amplitude $B_0(z)=B_0$ as a function of $z$ and the lack of fringe fields around the cavity apertures $z=0$ and $z=L$. The effect of both these non-ideal features are studied using particle tracking simulations in section \ref{sec:simulations}.

\begin{figure}[htp]
\centering
\includegraphics[width=0.5\linewidth]{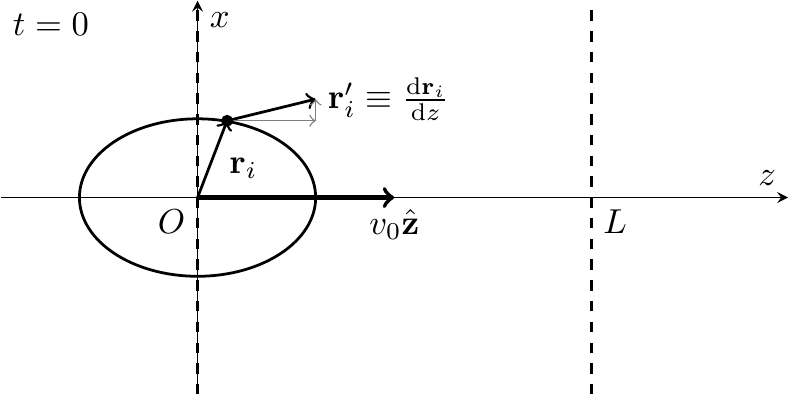}
\caption{The situation at time $t=0$ and the definition of $\mathbf{r}_i=(x_i,y_i,z_i)$ and $\mathbf{r}'_i= (x_i',y_i'z_i')$. At $t=0$ the center of the electron bunch enters the cavity at $z=0$ and the test particle is positioned at $\mathbf{r}_i$ and propagates with an angle $\mathbf{r}'_i\equiv\frac{\mathrm{d}\mathbf{r}_i}{\mathrm{d}z}$ with respect to the bunch center.}
\label{fig:electronbuncht0}
\end{figure}

The motion of a charged particle described by position vector $\mathbf{r}(t)=(x,y,z)$ and velocity vector $\mathbf{v}(t)=(v_x,v_y,v_z)\equiv (\dot{x},\dot{y},\dot{z})$ will be affected by the Lorentz force $\mathbf{F}=q(\mathbf{E}+\mathbf{v}\times \mathbf{B})$ as the particle travels through the cavity. This is described by the equations of motion
\begin{equation}\label{reom}
\begin{aligned}
\frac{\textrm{d}\mathbf{p}}{\textrm{d}t}&=\mathbf{F} = -q B_0
  \begin{pmatrix}
   v_z \cos(\phi_0+\omega t) \\
  0\\
   - v_x \cos(\phi_0+\omega t)+\omega x \sin(\phi_0+\omega t)
 \end{pmatrix}\\
\frac{\textrm{d}\mathbf{r}}{\textrm{d}t} &=  \mathbf{v}=\frac{\mathbf{p}}{\gamma m}, ~\text{with }~\gamma=\frac{1}{\sqrt{1-|\textbf{v}|^2/c^2}}.
\end{aligned}
\end{equation}
Now consider a 6D charged particle trace space distribution traveling along the $z$-axis with an average velocity $v_0\hat{\mathbf{z}}$. Figure \ref{fig:electronbuncht0} shows the moment $t=0$ at which the center of the distribution enters the cavity, which is indicated by the dashed lines. The black dot in figure \ref{fig:electronbuncht0} indicates a test particle that enters the cavity with trajectories
\begin{equation}\label{zeroth}
\begin{aligned}
\mathbf{v}(t)&\equiv v_0(x_i',y_i',1+z_i'),\\
\mathbf{r}(t)&\equiv(x_i,y_i,z_i)+v_0(x_i',y_i',1+z_i')t.
\end{aligned}
\end{equation}
Here we define ($x_i$,$y_i$,$z_i$) as the $t=0$ position of the test particle with respect to the bunch center, and we define $x_i'\equiv \frac{v_x}{v_0}$,$y_i'\equiv \frac{v_y}{v_0}$ and $z_i'\equiv \frac{\delta v_z}{v_0}$ as small deviations in propagation angles at $t=0$ in the frame of the traveling bunch center. Furthermore we define the 6D trace space coordinate $\mathbf{x}_i=(x_i,x_i',y_i,y_i',z_i,z_i')$ at $t=0$.

To obtain an approximate solution of the equations of motion \eqref{reom} for the test particle of which the original motion is described by expression \eqref{zeroth}, we use a perturbative approach based on the following two assumptions:
\begin{itemize}
\item{First, the charged particle gyrates only a small fraction of a full cyclotron orbit during one oscillation period of the EM field, hence
\begin{equation}\label{omegac}
\frac{\omega_c}{\omega}\ll 1,
\end{equation}
in which
\begin{equation}\label{cyclotron}
\omega_c\equiv\frac{q B_0}{\gamma_0 m}
\end{equation}
is the cyclotron frequency with $\gamma_0=\frac{1}{\sqrt{1-v_0^2/c^2}}$ the Lorentz factor of the incident beam. This ensures that the charged particles remain close to the $z$-axis and we can use the paraxial approximation throughout the paper.}
\item{Secondly, the distances over which the individual particles move with respect to the bunch center are small compared to the length scales of the collective motion of the bunch, i.e.
\begin{equation}\label{xiL}
x_i',y_i',z_i',\frac{x_i}{L},\frac{y_i}{L},\frac{z_i}{L}\ll \frac{\omega_c}{\omega}.
\end{equation}
}
\end{itemize}

\subsection{Transverse trajectories of the bunch center}
Based on assumption \eqref{xiL}, we first consider the motion of the bunch center, hence $\mathbf{x}_i=0$. Therefore we substitute $v_z=v_0$ and $v_x,x=0$ in equation \eqref{reom} and integrate from $t=0$ to $t$. The momentum of the bunch center calculated to first order in $\frac{\omega_c}{\omega}$ is then given by
\begin{equation}\label{pxtb}
\mathbf{p}^{(1)}(t)=\begin{pmatrix}\frac{B_0 q v_0}{\omega}\left(\sin{\phi_0}-\sin(\phi_0+\omega t)\right)\\0\\  \gamma_0 m v_0\end{pmatrix}.
\end{equation} 
Equation \eqref{pxtb} says that to first order, the bunch center is periodically deflected in the transverse direction while the longitudinal motion is unaffected. Because the transverse deflection is caused by the magnetic field, the kinetic energy remains unchanged, so that
\begin{equation}\label{vxtb}
\mathbf{v}^{(1)}(t)=\frac{\mathbf{p}^{(1)}(t)}{\gamma_0 m}=\begin{pmatrix}v_0 \frac{\omega_c}{\omega}\left(\sin{\phi_0}-\sin(\phi_0+\omega t)\right)\\0\\v_0\end{pmatrix}
\end{equation}
and
\begin{equation}\label{xt}
\mathbf{r}^{(1)}(t)=\begin{pmatrix}v_0 \frac{ \omega_c}{\omega^2}\left(-\cos{\phi_0}+\cos(\phi_0+\omega t)+\omega t \sin{\phi_0}\right)\\0\\v_0 t\end{pmatrix},
\end{equation}
in which we have substituted equation \eqref{cyclotron}. The deflection angle at which the pulse exits the cavity is given by
\begin{equation}\label{thetax}
\alpha=\frac{v_x}{v_z}=\frac{\omega_c}{\omega}\left(\sin{\phi_0}-\sin(\phi_0+\Lambda)\right),
\end{equation}
where we have introduced the dimensionless cavity length parameter
\begin{equation}
\Lambda\equiv\frac{\omega L_{}}{v_0}.
\end{equation}
When a small aperture of diameter $d$ is placed at a distance $l\gg d$ behind the cavity, centered along the cavity symmetry axis, only the electrons go through for which $|\alpha(\phi_0)|<\frac{d}{2l}\ll 1$, ignoring a small offset in $x$. This condition is satisfied for values of the RF phase close to
\begin{equation}\label{phi0}
\phi_0=\frac{\pi-\Lambda}{2}.
\end{equation}
In this regime, the deflection of the charged particles by the cavity can be considered as a linear function of the RF phase
\begin{equation}\label{alpha}
\alpha\approx\frac{\omega_c}{\omega}\left(\phi_0-\frac{\pi-\Lambda}{2}\right)\cdot 2 \sin\frac{\Lambda}{2},
\end{equation}
and the acceptance window of the slit in terms of the RF phase is given by
\begin{equation}\label{window}
-\frac{\omega d}{4\omega_c l\sin(\Lambda/2)}< \phi_0 - \frac{\pi-\Lambda}{2}<\frac{\omega d}{4\omega_c l\sin(\Lambda/2)}.
\end{equation}
The range of phases $\Delta \phi_0$ for which equation \eqref{window} is satisfied, determines the pulse length $\tau$ of the charged particle bunch behind the slit. Hence, an ideal continuous beam of charged particles is chopped up in temporally uniform pulses of pulse length
\begin{equation}\label{tau1}
\tau=\frac{\Delta \phi_0}{\omega}=\frac{\gamma_0 m d}{2 l q B_0\sin(\Lambda/2)}.
\end{equation}
Equation \eqref{tau1} shows that the pulse length of the resulting bunches for a given magnetic field amplitude can be minimized by choosing the cavity length parameter $\Lambda = \pi$, or $L=\frac{\pi v_0}{\omega}$. For such a cavity length the transit time of the charged particle traveling through the cavity equals half an oscillation period of the RF field. For 200 kV electrons traveling through a cavity with a typical resonance frequency of $\omega=2\pi\cdot 3$ GHz, this corresponds to a cavity length of $L_{}=35$ mm. In combination with a $d=10$ $\mu$m slit, positioned a distance $l=10$ cm behind the cavity and with a typical magnetic field strength of $B_0=3$ mT \cite{Lassise2012PhDthesis}, this results in $\tau=100$ fs pulses.

\begin{figure}[htp]
    \centering
        \includegraphics[width=\textwidth]{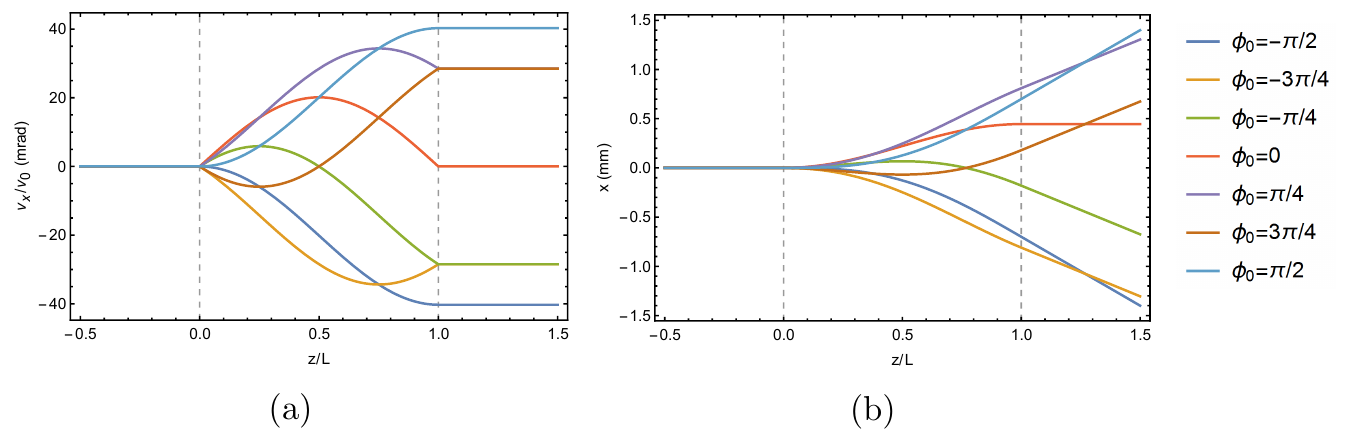}
    \caption{(a) Transverse deflection angle $v_x/v_0$ and (b) transverse position $x$ of the bunch center traveling through a TM$_{110}$ cavity as a function of normalized, longitudinal position $z/L$, for various values of the RF phase $\phi_0$. The electron energy is 200 keV, the cavity length is $L_{}=35$ mm, the angular frequency is $\omega = 2\pi \cdot 3$ GHz and the magnetic field amplitude is $B_0=3$ mT. The vertical, dashed lines indicate the extent of the cavity.}
    \label{fig:trajectories}
\end{figure}

The first order trajectories of the bunch center described by equations \eqref{vxtb} and \eqref{xt} for this situation are plotted in figure \ref{fig:trajectories} for various values of the RF phase $\phi_0$. The red curve in figure \ref{fig:trajectories} shows the charged particles that have experienced RF phase $\phi_0=0$, or more generally $\phi_0=(\pi-\Lambda)/2$. These particles exit the cavity with zero deflection angle, but with a small shift in transverse position $x$. To chop these particles of the beam, the chopping slit would have to be positioned slightly off-axis. However, for easy switching between pulsed mode and continuous mode, in practice the chopping slit is placed on the optical axis. As a result, the created electron pulses that go through the slit have a finite transverse momentum. Standard TEM deflection coils can be used to redirect the pulses back to the optical axis.

As a final remark, note that the spread in RF phase $\Delta\phi_0$ that defines the final pulses, also results in a spread of transverse momentum. Consequently, the transverse emittance is increased which reduces the focusability of the beam. In section \ref{sec:application} it is shown how this can be prevented.
\subsection{Longitudinal trajectories of the bunch center}
Due to the acquired motion in the $x$-direction, the bunch center now also starts to experience a Lorentz force in the $z$-direction. To calculate these second order effects, the first order expressions for $v_z$, $v_x$ and $x$ of equations \eqref{vxtb} and \eqref{xt} are substituted back into the equation of motion \eqref{reom}. The momentum of the bunch center calculated to second order in $\frac{\omega_c}{\omega}$ is then given by
\begin{equation}\label{pxt2}
\mathbf{p}^{(2)}(t)=\gamma_0 m v_0\begin{pmatrix}\frac{\omega_c}{\omega}\left(\sin{\phi_0}-\sin(\phi_0+\omega t)\right)\\0\\  1 + \frac{\omega_c^2}{\omega^2}\cos(\phi_0+\omega t)\left(-\cos{\phi_0}+\cos(\phi_0+\omega t)+\omega t \sin{\phi_0}\right)\end{pmatrix}.
\end{equation}
The work done by the electric field in the cavity results in a change in Lorentz factor
\begin{equation}\label{gamma}
\begin{aligned}
\Delta \gamma &=\frac{1}{m c^2}\int q \mathbf{E} \cdot \mathrm{d}\mathbf{r}\\
&=\gamma_0 \frac{ \omega_c^2}{\omega^2}  \frac{v_0^2}{c^2} \left[1-  \cos{ \omega t} +   \omega t \cos(\phi_0 + \omega t) \sin{\phi_0}  +\frac{ \cos(2 (\phi_0 + \omega t))-\cos{2 \phi_0}}{4}\right],
\end{aligned}
\end{equation}
in which we also have substituted the expression for $x$ of equation \eqref{xt}. With this change in Lorentz factor $\gamma$ the second order longitudinal velocity and position are given by
\begin{equation}\label{second2}
\begin{aligned}
v_z^{(2)}(t)&=\frac{ p_z^{(2)}(t)}{m (\gamma_0+\Delta \gamma)}\approx \frac{ p_z^{(2)}(t)}{\gamma_0 m}\left(1-\frac{\Delta \gamma}{\gamma_0}\right)\\
&=v_0 + v_0\frac{ \omega_c^2}{  \omega^2}\left\{ \cos(\phi_0 + \omega t) \left(-\cos\phi_0 + \cos(\phi_0 + \omega t) +  \omega t \sin\phi_0\right)\right.\\
 &\left.~- \frac{v_0^2}{c^2} \left[1-  \cos{ \omega t} +   \omega t \cos(\phi_0 +  \omega t) \sin{\phi_0}  -\frac{\cos{2 \phi_0}-  \cos(2 (\phi_0 + \omega t))}{4}\right]\right\}
\end{aligned}
\end{equation}
and
\begin{equation}\label{second3}
\begin{aligned}
z^{(2)}(t)&=v_0 t + \frac{v_0 \omega_c^2}{ \omega^3} \left\{\frac{\omega t}{2} - \sin \omega t+   \omega t \sin\phi_0 \sin(\phi_0 + \omega t) +\frac{\sin(2 (\phi_0 +  \omega t)) -\sin2 \phi_0}{4}\right. \\
    &~\left.-\frac{v_0^2}{c^2} \left[ \omega t +  \sin\phi_0(-\cos\phi_0 + \cos(\phi_0 +  \omega t) +  \omega t \sin(\phi_0 +  \omega t))\right.\right.\\
    &~\left.\left.-\frac{ \omega t \cos2 \phi_0+ (1 - \cos(2 \phi_0 +  \omega t)) \sin \omega t}{4}\right] \right\}
\end{aligned}
\end{equation}
in which we have assumed $\frac{\Delta \gamma}{\gamma_0}\ll 1$, and have omitted terms proportional to $\left(\frac{\omega_c}{\omega}\right)^3$ and higher, based on assumption \eqref{omegac}.

\begin{figure}[htp]
    \centering
        \includegraphics[width=\textwidth]{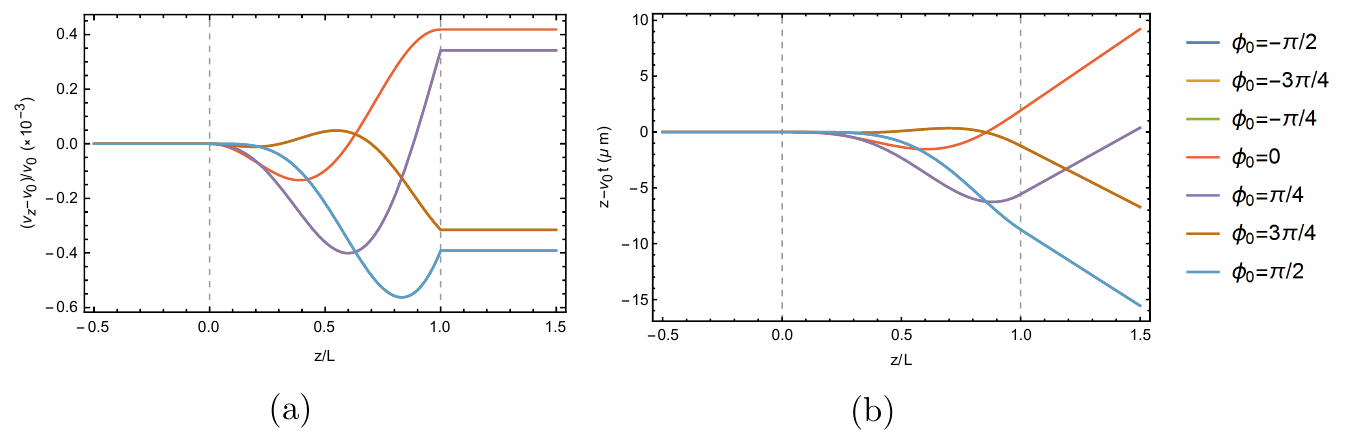}
    \caption{(a) Relative change in longitudinal velocity $(v_z-v_0)/v_0$ and (b) resulting change in longitudinal position $z-v_0 t$ of the bunch center traveling through a TM$_{110}$ cavity with respect to a frame traveling with its initial velocity $v_0 \hat{\mathbf{z}}$ as a function of normalized, longitudinal position $z/L$ , for various values of the RF phase $\phi_0$. The electron energy is 200 keV, the cavity length is optimized to $L_{}=35$ mm, the angular frequency is $\omega = 2\pi \cdot 3$ GHz and the magnetic field amplitude is $B_0=3$ mT. The vertical, dashed lines indicate the extent of the cavity.}
    \label{fig:trajectorieslong}
\end{figure}

Figure \ref{fig:trajectorieslong} shows the longitudinal trajectories of equations \eqref{second2} and \eqref{second3} relative to a co-moving frame traveling with the initial velocity of the bunch $v_0\hat{\mathbf{z}}$, for various values of the RF phase $\phi_0$. In other words, figure \ref{fig:trajectorieslong}a shows the relative change in longitudinal velocity of the bunch $(v_z-v_0)/v_0$ as a function of longitudinal position $z$ in the cavity. Figure \ref{fig:trajectorieslong}b shows the resulting deviation in longitudinal position relative to the moving frame. Again, the red curve shows the charged particles that have experienced RF phase $\phi_0=0$, or more generally $\phi_0=(\pi-\Lambda)/2$. Figure \ref{fig:trajectorieslong} shows that these particles are first decelerated and subsequently accelerated. The spread in RF phase $\Delta \phi_0$ experienced in the bunch results in an increased energy spread. This is also addressed in section \ref{sec:application}.

\section{Optical transfer matrix}\label{sec:matrix}
To derive the optical transfer matrix of a RF cavity in TM$_{110}$ mode we follow the same perturbative approach as in section \ref{sec:trajectories} to calculate the trajectories of particles with $\mathbf{x}_i\neq0$, see figure \ref{fig:electronbuncht0}. Because of assumptions \eqref{omegac} and \eqref{xiL}, the additional coupling between the $\mathbf{x}_i$ terms and the fields in the cavity is regarded as a second order effect. Therefore, the first step is to solve the equation of motion for the test particle described by initial trajectories \eqref{zeroth}, while substituting $\mathbf{x}_i=0$ in the expression for the Lorentz force. Note that now the equations of motions must be integrated from
\begin{equation}\label{tbte}
t_b\equiv\frac{-z_i}{v_0(1+z_i')}
 \end{equation}
to $t$, because the test particle no longer enters the cavity at $t=0$. With this, the first order trajectories of the test particle are given by
\begin{equation}\label{vxt}
\begin{aligned}
\mathbf{v}^{(1)}(t)&=\begin{pmatrix}v_0 x_i'+v_0 \frac{\omega_c}{\omega}\left(\sin{(\phi_0+\omega t_b)}-\sin(\phi_0+\omega t)\right)\\v_0 y_i'\\v_0 (1+z_i')\end{pmatrix} \textrm{~and}\\
\mathbf{r}^{(1)}(t)&=\begin{pmatrix}x_i+v_0 x_i' t+ v_0 \frac{ \omega_c}{\omega^2}\left(\cos(\phi_0+\omega t)-\cos(\phi_0+\omega t_b)+\omega(t-t_b)\sin(\phi_0+\omega t_b)\right)\\y_i+v_0 y_i' t\\z_i+v_0(1+z_i')t\end{pmatrix}.
\end{aligned}
\end{equation}
Subsequently, the first order expressions for $v_z$, $v_x$ and $x$ of equation \eqref{vxt} are substituted back in equation \eqref{reom} to calculate the second order trajectories. This results in lengthy expressions for $\mathbf{v}^{(2)}(t)$ and $\mathbf{r}^{(2)}(t)$ with many cross-terms of the various initial particle coordinates $\mathbf{x}_i$. However because of assumptions \eqref{omegac} and \eqref{xiL}, only effects that are linearly dependent on initial particle coordinates $\mathbf{x}_i$ and up to second order in $\frac{\omega_c}{\omega}$ are taken into account. This allows the definition of the optical transfer matrix $\underline{M}_{cav}$ via
  \begin{equation}
 \mathbf{x}_f=\underline{M}_{cav}\mathbf{x}_i,
 \end{equation}
as the linear transformation that maps the initial 6D trace space coordinate $\mathbf{x}_i$ at $t=0$ onto the final 6D trace space coordinate $\mathbf{x}_f=(x_f,x_f',y_f,y_f',z_f,z_f')$, defined at the time $t_e$ at which the test particle exits the cavity:
\begin{equation}
t_e=\frac{L-z_i}{v_0(1+z_i')}.
\end{equation}
More specifically, we define
\begin{equation}\label{xf}
\begin{pmatrix} x_f\\y_f\\z_f\end{pmatrix} \equiv\begin{pmatrix}x^{(2)}(t_e)\\y^{(2)}(t_e)\\z^{(2)}(t_e)-v_0 t_e\end{pmatrix} \textrm{~and~}
\begin{pmatrix}x_f'\\y_f'\\z_f'\end{pmatrix}\equiv \frac{1}{v_0}\begin{pmatrix} v_x^{(2)}(t_e)\\ v_y^{(2)}(t_e)\\ v_z^{(2)}(t_e)-v_0\end{pmatrix}
 \end{equation}
as the position and propagation angle of the test particle at time $t_e$ in the frame of the traveling bunch center. Note that in this step we have assumed the longitudinal velocity of the charged particle remains constant while traversing the cavity.

After calculating the second order trajectories for the test particle, evaluating them at time $t=t_e$ and expanding them to first order in $\mathbf{x}_i$, we obtain the optical transfer matrix for a TM$_{110}$ cavity
\begin{equation}\underline{M}_{cav}=\begin{pmatrix}
 1      & L_{}   & 0         & 0     & C_1       & C_{2}\\
 0      & 1         & 0         & 0     & C_{3}     & C_{4}\\
 0      & 0         & 1         & L_{} & 0         & 0 \\
 0      & 0         & 0         & 1       & 0         & 0\\
 C_{5}  & C_{6}     & 0         & 0        & 1+C_{9}   &L_{}+ C_{10}\\
 C_{7}  & C_{8}     & 0         & 0        & C_{11}    & 1+C_{12} \end{pmatrix},\end{equation}
in which the cavity constants $C_{1}$ through $C_{12}$ are given by
\begin{equation}C_{1}=-\frac{\omega_c}{\omega} \left(\Lambda \cos{\phi_0} + \sin{\phi_0} - \sin{\left(\Lambda + \phi_0\right)}\right),\end{equation}
\begin{equation}C_{2}=\frac{L_{}}{\Lambda}\frac{ \omega_c}{\omega} \left(\cos(\Lambda+\phi_0)-\cos{\phi_0}+\Lambda \sin(\Lambda+\phi_0)\right),\end{equation}
\begin{equation}C_{3}=-\frac{\Lambda}{L_{}}\frac{ \omega_c}{\omega} \left(\cos{\phi_0} - \cos{\left(\Lambda + \phi_0\right)}\right),\end{equation}
\begin{equation}C_{4}=\frac{\omega_c}{\omega}\left(\Lambda\cos(\Lambda+\phi_0)+\sin \phi_0 -\sin(\Lambda + \phi_0)\right),\end{equation}
\begin{equation}C_{5}=\frac{C_1}{\gamma_0^2},\end{equation}
\begin{equation}C_{6}=\frac{L_{}}{\Lambda}\frac{ \omega_c}{\omega} \left[(1-2\beta_0^2)\left(\cos(\Lambda+\phi_0)-\cos{\phi_0}\right)-\beta_0^2 \Lambda \sin{\phi_0}+\frac{\Lambda}{\gamma_0^2}\sin(\Lambda+\phi_0)\right],\end{equation}
\begin{equation}C_{7}=\frac{C_3}{\gamma_0^2},\end{equation}
\begin{equation}C_{8}=\frac{\omega_c}{\omega}\left(\frac{\Lambda}{\gamma_0^2}\cos(\Lambda+\phi_0)+\beta_0^2(\sin(\Lambda + \phi_0)-\sin \phi_0)\right),\end{equation}
\begin{equation}\begin{aligned}
C_{9}=&~\frac{\omega_c^2}{\omega^2} \left[\left(\frac{1}{2} - \frac{5 }{4}\beta_0^2\right) \cos( 2 \phi_0) + \left(-\frac{1}{2}+\frac{\beta_0^2}{4}\right) \cos(2 (\Lambda + \phi_0))\right.\\
 &\left.+\beta_0^2 ( \cos(\Lambda + 2 \phi_0 )+ \frac{\Lambda}{2} \sin(2 \phi_0)) - \frac{\Lambda}{\gamma_0^{2}} \sin(\Lambda + 2 \phi_0)\right],\end{aligned}\end{equation}
\begin{equation}\begin{aligned}
C_{10}=&~L\frac{\omega_c^2}{\omega^2}\left[ \frac{1}{2}\left(1-4\beta_0^2\right)\cos \Lambda + \frac{\beta_0^2}{4} \cos 2\phi_0+\frac{1}{2}\left(1-\frac{\beta_0^2}{2}\right)\cos(2(\Lambda+\phi_0))\right.\\
&\left.+\frac{1}{2}\left(1+\beta_0^2\right)\left[\cos(\Lambda+2\phi_0)-1\right]+\left(\frac{3\beta_0^2}{\Lambda}+\frac{\Lambda}{2\gamma_0^2}\right)\sin \Lambda \right.\\
&\left.+\frac{5\beta_0^2}{4\Lambda}\sin 2\phi_0 -\frac{\beta_0^2}{4\Lambda}\sin(2(\Lambda+\phi_0))\right],\end{aligned}\end{equation}
\begin{equation}\begin{aligned}
C_{11}=&~\frac{\omega_c^2}{\omega^2}\frac{\Lambda}{L}\left[ \sin(2 (\Lambda + \phi_0)) -\sin(\Lambda +2\phi_0)-\left(\frac{\Lambda}{\gamma_0^{2}} + \beta_0^2 \sin\Lambda\right)\cos(\Lambda + 2 \phi_0)  \right]\end{aligned}\end{equation}
and
\begin{equation}\begin{aligned}
C_{12}=&~\frac{\omega_c^2}{\omega^2}\left[\Lambda\left\{ \sin(2 (\Lambda + \phi_0)) -  \sin(\Lambda + 2 \phi_0)+\Lambda \sin \phi_0  \sin(\Lambda + \phi_0)\right\}\right.\\
 &\left.- \beta_0^2  \left(   \cos( \Lambda + \phi_0)-\cos \phi_0 + \Lambda \sin \phi_0 \right)\Lambda \sin(\Lambda + \phi_0)\right.\\
 &\left.-2\beta_0^2 \left(1 - \cos \Lambda+  \Lambda \cos(\Lambda + \phi_0) \sin \phi_0  + \frac{1}{4}\left[\cos(2 (\Lambda + \phi_0))-\cos 2 \phi_0 \right] \right)\right],\end{aligned}\end{equation}
with
\begin{equation}
\beta_0\equiv\frac{v_0}{c}
\end{equation}
the normalized initial speed of the bunch center.

Note that cavity constants $C_1$-$C_8$ are proportional to the magnetic field amplitude and therefore scale linearly with $\frac{\omega_c}{\omega}$. The cavity constants $C_9$-$C_{12}$ represent second order effects scaling with $\left(\frac{\omega_c}{\omega}\right)^2$. In the limit where the fields are turned off ($\omega_c\rightarrow 0$), all the cavity constants vanish and the matrix $\underline{M}_{cav}$ simply reduces to a drift over a distance $L_{}$. The determinant of the transfer matrix is unity to first order in $\frac{\omega_c}{\omega}$, which means that the 6D trace space density of a charged particle distribution is conserved during cavity transit. Furthermore, note that the transfer matrix $\underline{M}_{cav}$ simplifies significantly for an optimized cavity length of $L_{}=\frac{\pi v_0}{\omega}$, i.e. $\Lambda=\pi$.

\section{Courant-Snyder trace-space transformation}\label{sec:CS}

\begin{figure}[htp]
\centering
\includegraphics[width=0.5\linewidth]{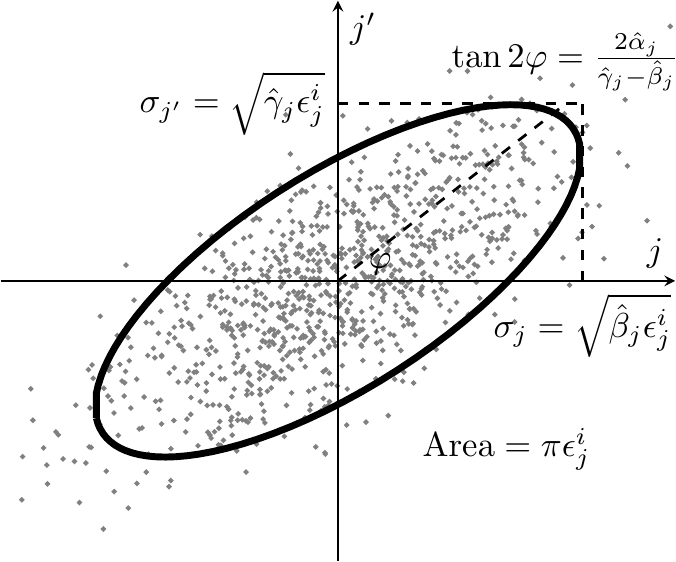}
\caption{Trace space distribution of charged particles projected on the $j,j'$-plane ($j=x,y,z$) with an indicative, elliptical contour of equal trace space density, described by equation \eqref{twiss}. The area occupied by the ellipse is given by $\pi \epsilon^i_j$. All calculations are done for rms quantities for $\sigma_j$, $\sigma_j'$ and $\epsilon_j^i$. }
\label{fig:ellipse}
\end{figure}

Now we have the optical transfer matrix of the TM$_{110}$ cavity, we use the Courant-Snyder formalism \cite{Courant1958} to calculate the propagation of an entire 6D trace-space distribution of charged particles. For this purpose, we describe the distribution in terms of its rms ellipsoidal contours in trace space, of which the projections on the $(j,j')$ planes ($j=x,y,z$) are given by the ellipses
\begin{equation}\label{twiss}
\epsilon^i_j=\hat{\gamma}_j j^2+2\hat{\alpha}_j j j'+\hat{\beta}_j j^2 \textrm{~with~}\hat{\beta}_j\hat{\gamma}_j-\hat{\alpha}_j^2=1.
\end{equation}
Figure \ref{fig:ellipse} shows how the Courant-Snyder parameters $\hat{\alpha}_j,\hat{\beta}_j,\hat{\gamma}_j$ are related to initial beam properties such as rms radius $\sigma_j$ and rms angular divergence $\sigma_{j'}$. The rms area of the ellipse is given by $\pi \epsilon^i_j$, where $\epsilon^i_j$ is the (initial) projected rms emittance in the $j,j'$-plane.

We can write equations \eqref{twiss} in matrix notation
\begin{equation}
\mathbf{x}^T \underline{A}^{-1} \mathbf{x}=1,
 \label{ellipseeq}
 \end{equation}
with $\underline{A}$ the real, positive definite, $6\times6$ \emph{beam matrix} at $t=0$ given by
\begin{equation}\underline{A}=
\begin{pmatrix}
 \underline{A}_{xx'} & \underline{0} & \underline{0} \\
 \underline{0} & \underline{A}_{y y'} & \underline{0} \\
  \underline{0} & \underline{0} & \underline{A}_{z z'}
 \end{pmatrix} \textrm{~with~}\underline{A}_{j j'}=\epsilon^i_j
\begin{pmatrix}
 \hat{\beta}_j &-\hat{\alpha}_j\\
 -\hat{\alpha}_j &\hat{\gamma}_j
  \end{pmatrix}.
 \end{equation}
Here we assume no correlations between spatial degrees of freedom. Note that $\epsilon^i_j=\sqrt{\det(\underline{A}_{jj'})}$ because of equation \eqref{twiss}. We can now propagate this charged particle distribution through an ideal TM$_{110}$ cavity. The beam matrix that describes the distribution at the exit of the cavity $t=t_e$ is given by
\begin{equation}\label{Af}
\underline{B}=\underline{M}_{cav} \underline{A} \underline{M}_{cav}^T\equiv
\begin{pmatrix}
 \underline{B}_{xx'} & \underline{0} & \textrm{cross terms} \\
 \underline{0} & \underline{B}_{yy'} & \underline{0} \\
  \textrm{cross terms}& \underline{0} & \underline{B}_{zz'}
 \end{pmatrix}
.
 \end{equation}
The $x,z$-correlations are introduced by the non-zero, off-diagonal matrix elements of $\underline{M}_{cav}$ and will cause an exchange between the transverse and longitudinal emittance, hence energy spread. Both will decrease the focusability of the beam.

\begin{figure}[htp]
\centering
\includegraphics[width=0.5\linewidth]{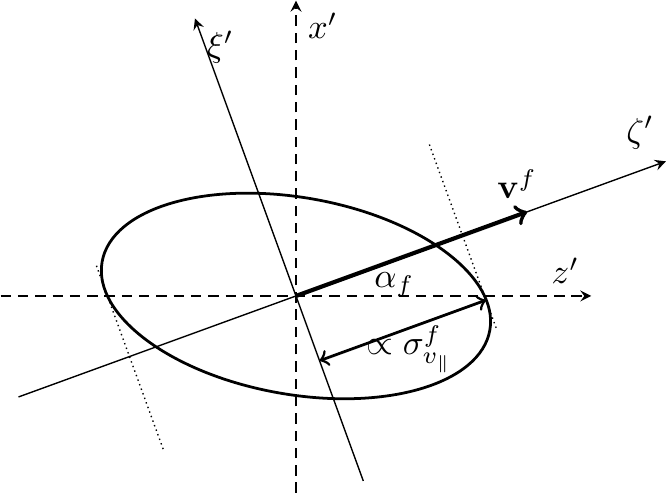}
\caption{The 2D beam ellipse that describes the final trace space distribution projected on the $(x',z')$-plane, as described by beam matrix $\underline{B}_{x'z'}$ in equation \eqref{Bxz}, in the $(x',z')$ coordinate system and the rotated $(\xi',\zeta')$ coordinate system. The diagonal matrix elements in the latter represent the velocity spread in the direction parallel en perpendicular to the velocity of the bunch.}
\label{fig:rotation}
\end{figure}

The final normalized transverse emittance is calculated by
\begin{equation}\label{emfinal}
\epsilon^f_{n,x}=\beta_0\gamma_0\sqrt{\mathrm{det}(\underline{B}_{xx'})}
\end{equation}
while the rms energy spread of the beam after cavity transit is given by
\begin{equation}
\sigma_U^f=\gamma_0^3 m v_0 \sigma_{v_\parallel}^f,
\end{equation}
with $\sigma_{v_\parallel}^f$ the rms spread in velocity along the propagation vector $\mathbf{v}^f$ of the deflected beam. To find $\sigma_{v_\parallel}^f$ we consider the 2D beam ellipse that describes the final trace space distribution projected on the $(x',z')$-plane
\begin{equation}
\underline{B}_{x'z'} \equiv
\begin{pmatrix}
  B_{(2,2)} & B_{(2,6)} \\
  B_{(6,2)} & B_{(6,6)}
 \end{pmatrix}.
 \label{Bxz}
 \end{equation}
The diagonal matrix elements $B_{(2,2)}$ and $B_{(6,6)}$ are related to the velocity spread of the bunch in the $x$ and $z$ directions, via $\sigma^f_{v_x}=v_0 \sqrt{B_{(2,2)}}$ and $\sigma^f_{v_z}=v_0 \sqrt{B_{(6,6)}}$ respectively. Next, we consider this beam matrix in the ($\xi',\zeta'$)-coordinate system as illustrated in figure \ref{fig:rotation}, which is rotated with respect to the $(x',z')$-coordinate system about the final propagation angle at which the bunch center exits the cavity
 \begin{equation}
 \begin{aligned}
\alpha_f\equiv &~\frac{v^f_x}{v^f_z}= \frac{x_f'}{1+z_f'}\\
= &~\frac{\omega_c}{\omega} (\sin \phi_0 -\sin(\Lambda + \phi_0))\Biggm/\left\{1+\frac{\omega_c^2}{\omega^2}\left[\cos(\Lambda+\phi_0)\left(\cos(\Lambda+\phi_0)+\frac{\Lambda}{2}\sin \phi_0-\cos\phi_0\right)\right.\right.\\
&\left.\left.-\frac{v_0^2}{c^2}\left(1-\cos \Lambda +\Lambda\cos(\Lambda+\phi_0)\sin\phi_0-\frac{\sin\Lambda\sin(\Lambda+2\phi_0)}{2}\right)\right]\right\}.
\end{aligned} \end{equation}
In this coordinate system, the $\xi'$ and $\zeta'$ axes are perpendicular and parallel to the final velocity vector $\mathbf{v}^f$ of the bunch, respectively.
The final beam matrix in the rotated coordinate system is given by
\begin{equation}
 \underline{B}^{rot}_{\xi'\zeta'}\equiv \underline{M}_{rot}\underline{B}_{x'z'}\underline{M}_{rot}^T\textrm{~with~}
 \underline{M}_{rot}\equiv
 \begin{pmatrix}
  1-\alpha^2_f & -\alpha_f \\
  \alpha_f & 1-\alpha^2_f
 \end{pmatrix},
 \end{equation}
of which the diagonal matrix elements $B^{rot}_{(2,2)}$ and $B^{rot}_{(6,6)}$ are related to the velocity spread parallel and perpendicular to $\mathbf{v}^f$, via $\sigma^f_{v_\perp}=v_0 \sqrt{B^{rot}_{(2,2)}}$ and $\sigma^f_{v_\parallel}=v_0 \sqrt{B^{rot}_{(6,6)}}$ respectively. Now the final energy spread of the bunch after propagating through the cavity is given by
\begin{equation}\label{esfinal}
\sigma_U^f=\gamma_0^3 m v_0 \sigma_{v_\parallel}^f=\gamma_0^3 m v_0^2 \sqrt{B^{rot}_{(2,2)}}.
\end{equation}

So, using the Courant-Snyder formalism, we can derive analytical expressions for the final normalized transverse emittance (equation \eqref{emfinal}) and final energy spread (equation \eqref{esfinal}) of the beam after traversing an ideal TM$_{110}$ cavity. Moreover these expressions are derived as function of the initial transverse emittance and energy spread of the incident beam. To our knowledge, this is not possible in any other way.

\section{Application: focused beam in a 200 kV TEM}\label{sec:application}
Now we apply the Courant-Snyder model to the special case of a focused beam in a 200 kV TEM. More specific, we calculate the increase in normalized transverse emittance and energy spread of a 200 kV 6D Gaussian charged particle distribution with a finite initial geometrical emittance and energy spread, focused to a crossover inside a TM$_{110}$ cavity. We are aware that Gaussian distributions are not realistic in electron microscopes, but they result in easy calculations for the rms quantities, required for the Courant-Snyder model. Furthermore, the functional dependencies in the final expressions are independent of the shape of distribution, the only difference is a proportionality factor. In section \ref{sec:simulations}, we use charged particle tracking simulations to calculate actual numbers. Furthermore, the Courant-Snyder model obliges us to choose a finite initial pulse length, although in the experiment the initial beam is continuous. However, we are not interested in the electrons that are not part of the final (chopped) pulse. So by choosing an initial pulse length equal to the expected final pulse length we simply leave out the electrons of which we already know they will collide into the chopping aperture. In section \ref{sec:simulations}, we use charged particle tracking simulations to test the validity of this approach.

\begin{figure}[htp]
\centering
\includegraphics[width=0.5\linewidth]{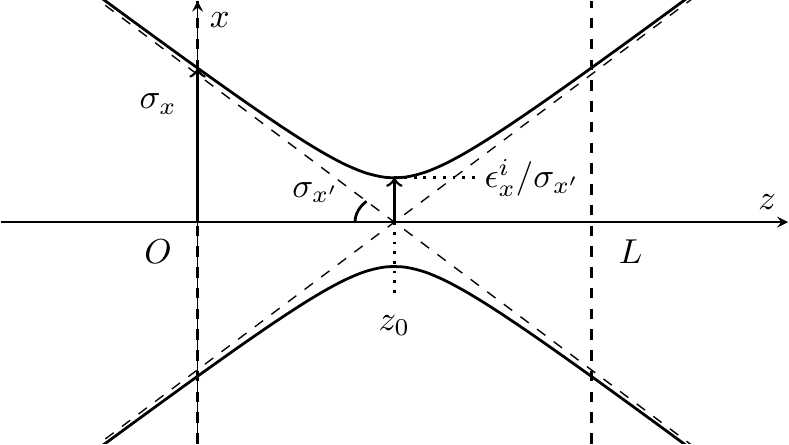}
\caption{Schematic illustration of a charged particle beam with a crossover at position $z=z_0$ inside a TM$_{110}$ cavity of length $L$, indicated by the vertical dashed lines. The rms divergence angle of the beam is $\sigma_{x'}$ and the rms geometrical emittance is $\epsilon_x^i$. Therefore the rms radius of the beam in the crossover $z=z_0$ is $\epsilon_x^i/\sigma_{x'}$ and the rms beam radius at the cavity entrance $z=0$ is given by $\sigma_x=\sqrt{\left(\frac{\epsilon_x^i}{\sigma_{x'}}\right)^2+\left(\sigma_{x'} z_0\right)^2}$.}
\label{fig:focusedbeam}
\end{figure}

Figure \ref{fig:focusedbeam} shows an electron beam with initial geometrical emittance $\epsilon_x^i$ that is focused to a crossover at $z=z_0$ with rms divergence angle $\sigma_{x'}$. Therefore the rms beam radius at $z=0$ is given by $\sigma_x=\sqrt{\left(\frac{\epsilon_x^i}{\sigma_{x'}}\right)^2+\left(\sigma_{x'} z_0\right)^2}$. Furthermore the distribution has an initial rms pulse duration $\sigma_t$, initial rms energy spread $\sigma_U^i$ and no initial chirp.

The beam matrix $\underline{A}$ that describes this distribution at $t=0$ is defined by the Courant-Snyder parameters: $\hat{\beta}_{x}=\frac{\sigma_x^2}{\epsilon_{x}^i}$, $\hat{\gamma}_{x}=\frac{\sigma_{x'}^2}{\epsilon_{x}^i}$, $\hat{\alpha}_{x}=\sqrt{\hat{\beta}_{x}\hat{\gamma}_{x} -1}$, $\epsilon_z^i\hat{\beta}_z=\left(v_0 \sigma_t\right)^2$, $\epsilon_z^i\hat{\gamma}_z=\left(\frac{\sigma_U^i}{\gamma_0^3 m v_0^2}\right)^2$ and $\hat{\alpha}_z=0$. The beam matrix $\underline{B}$ that describes the beam at the exit of the cavity may be calculated using equation \eqref{Af}. Then using equations \eqref{emfinal} and \eqref{esfinal} we can calculate the final normalized transverse emittance and energy spread of the beam.
\begin{figure}[htp]
 \centering
        \includegraphics[width=\textwidth]{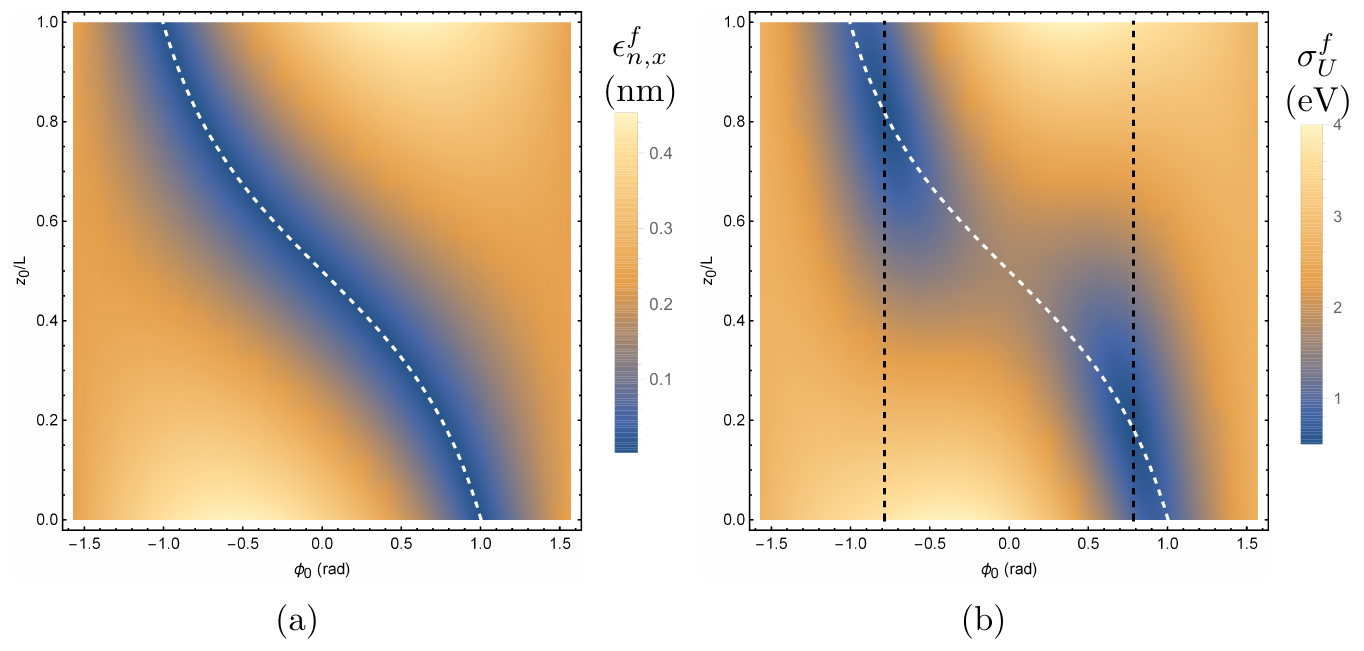}
        \label{fig:emesgrowth}
    \caption{Final normalized transverse emittance $\Delta \epsilon_{n,x}^f$ and energy spread $\sigma_U^f$ of a 200 keV electron beam with no chirp, focused in an ideal TM$_{110}$ cavity as a function of RF phase $\phi_0$ and focal point $z_0$. Beam parameters: $\epsilon_{n,x}=3$ pm rad, $\sigma_{x'} = 0.15$ mrad, $\sigma=200$ fs and $\sigma_U^i=0.5$ eV. Cavity parameters: $\omega = 2\pi \cdot 3$ GHz, $B_0=3$ mT and an optimized length of $L=35$ mm. The white dashed curve is described by equation \eqref{emzero}, the black dashed lines are described by equation \eqref{taucontribution}.}\label{fig:emesgrowth}
\end{figure}

Figure \ref{fig:emesgrowth} shows (a) the final normalized transverse emittance $\epsilon^f_{n,x}$ and (b) the final energy spread $\sigma_U^f$ of the 200 keV electron beam directly behind the TM$_{110}$ cavity as a function of the cavity RF phase $\phi_0$ and the position of the crossover $z_0$ for typical beam parameters of a 200 kV (pulsed) TEM: $\epsilon_{n,x}=3$ pm rad, $\sigma_{x'} = 0.15$ mrad, $\sigma_t=200$ fs, $\sigma_U^i=0.5$ eV; and cavity parameters: $\omega = 2\pi \cdot 3$ GHz, $B_0=3$ mT and $L=35$ mm. The figure shows a dark blue region in $(\phi_0,z_0)$ parameter space for which both quantities hardly increase and the quality of the incident beam is maintained. This will be investigated further in the next sections.

\subsection{Transverse emittance}
The expression for the final transverse normalized emittance $\epsilon_{n,x}^f$ that is calculated by equation \eqref{emfinal} and is plotted in figure \ref{fig:emesgrowth}a, is lengthy and does not provide much insight. However, if we assume an ideal incident electron beam: hence no initial transverse emittance $\epsilon_{n,x}^i=0$ and energy spread $\sigma_U^i=0$, equation \eqref{emfinal} reduces to a closed, analytic expression for the final, normalized transverse emittance:
\begin{equation}\label{emgr}
\boxed{
\epsilon_{n,x}^f=\beta_0\gamma_0\sigma_{x'} \sigma_t \frac{\omega_c}{\omega}
\left| \frac{z_0}{L_{}} \Lambda \cos{\phi_0} - \left(\frac{z_0}{L_{}}-1\right)\Lambda \cos{(\Lambda+\phi_0)}+\sin{\phi_0}-\sin{(\Lambda+\phi_0)}  \right|.
}
\end{equation}
Note that the right-hand side of equation \eqref{emgr} is zero for
\begin{equation}\label{conjugate}
\frac{z_0}{L_{}}=\frac{\Lambda\cos{(\Lambda+\phi_0)}-\sin{(\Lambda+\phi_0)}+\sin{\phi_0}}{\Lambda(\cos{(\Lambda+\phi_0)}-\cos{\phi_0})},
\end{equation}
and for a cavity with optimized cavity length $\Lambda=\pi$, this reduces to
\begin{equation}\label{emzero}
\frac{z_0}{L_{}}=\frac{1}{2}-\frac{\tan{\phi_0}}{\pi}.
\end{equation}
Equation \eqref{emzero} describes the white dashed curve in figure \ref{fig:emesgrowth}a. It describes a region in parameter space where propagation of an ideal beam through an RF cavity in TM$_{110}$ mode results in \emph{zero increase in transverse emittance}. In the situation of entrance phase $\phi_0=\frac{1}{2}(\pi-\Lambda)$, for which the shortest pulses are obtained, equation \eqref{conjugate} reduces to $\frac{z_0}{L}=\frac{1}{2}$, hence focusing the electron beam in the center of the cavity.

\begin{figure}[htp]
    \centering
        \includegraphics[width=\textwidth]{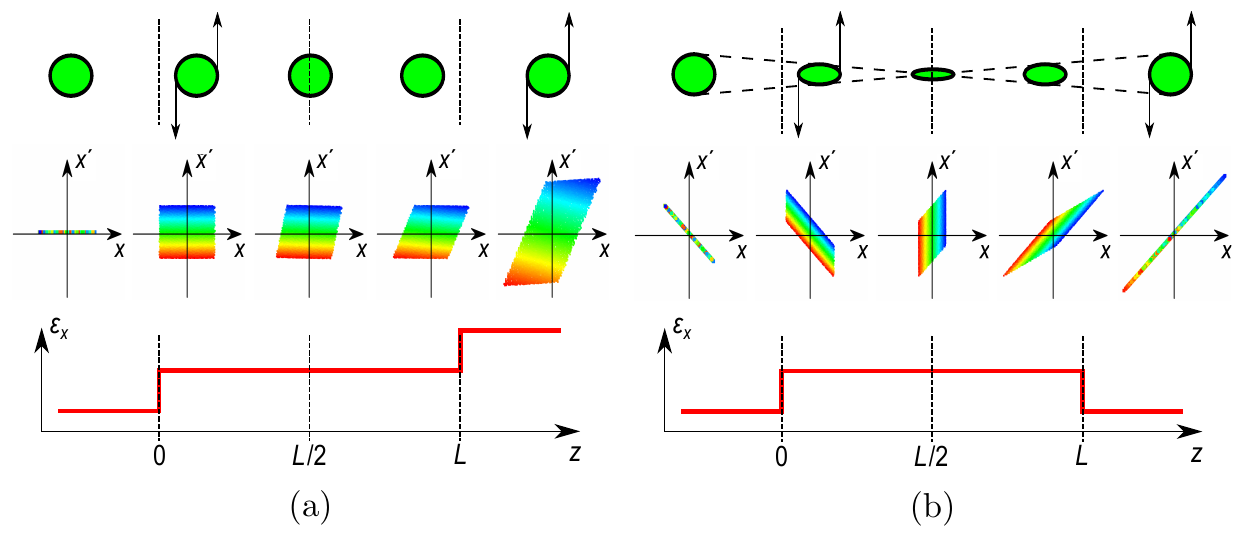}
    \caption{Conjugate blanking: Real ($x,z$)-space (green), ($x',x$) phase-space (color coded for time) and transverse emittance $\epsilon_{x}$ (red curve) as a function of longitudinal coordinate $z$ of (a) a collimated and (b) a focused beam propagating through a TM$_{110}$ cavity. Color coding: Blue = front, red = back. Note that the transverse emittance is proportional to the ($x,x'$) trace-space area.}
\label{fig:emgrexplained}
\end{figure}

Figure \ref{fig:emgrexplained} explains the principle of \emph{conjugate blanking} for $\phi_0\approx(\pi-\Lambda)/2$ in more detail. It shows the real ($x,z$)-space (top), ($x',x$) phase-space (middle) and transverse emittance $\epsilon_{x}$ (bottom) as a function of longitudinal coordinate $z$ for both a collimated and a focused beam propagating through a TM$_{110}$ cavity. The color coding shows the correlation with time, for which blue indicates the front and red indicates the back of the pulse. Furthermore, note that the transverse emittance is proportional to the ($x,x'$) trace-space area.

When a bunch of charged particles enters the cavity at $z=0$, the particles suddenly feel a force in the transverse direction, which results in a transverse deflection. This force varies with time, hence the front of the pulse feels a slightly different force than the back of the pulse. This results in an angular spread and therefore an increase in transverse emittance. The fields in the cavity are homogeneous along the $z$-axis, so during transit through the cavity, all the particles in the bunch experience the same forces and the emittance is unaffected. However, when the bunch arrives at the exit aperture, again there is a sharp step in the experienced Lorentz force, that changes in amplitude while the bunch travels past this gradient. Figure \ref{fig:emgrexplained}a shows that for a collimated beam, this results in a second increase in angular spread, hence a second emittance growth. Note that the increase in emittance at the apertures is not an effect of fringe fields.

However, by focusing the beam in the center of the cavity (figure \ref{fig:emgrexplained}b) the angular spread that is obtained at the entrance aperture, can be canceled by the forces at the exit aperture. This is seen best in the ($x,x'$) trace space plot. By focusing, an additional correlation is applied between $x$ and $x'$. During passage through the cavity, the ($x,x'$) trace space distribution is sheared parallel to the $x$-axis in such a way that at the exit of the cavity the trace space distribution is collapsed onto a line, thus canceling the emittance growth at the entrance of the cavity. This is called conjugate blanking.

For an on-axis slit, hence $\phi_0=(\pi-\Lambda)/2$, the focus point for conjugate blanking lies exactly in the center of the cavity ($z_0=L/2$). However, for any other phase than $\phi_0=(\pi-\Lambda)/2$ the experienced Lorentz forces at the entrance and exit of the cavity are not symmetric. By focusing the beam at a different position given by equation \eqref{conjugate} the emittance growth at the entrance of the cavity can still be fully canceled at the exit.

\subsection{Pulse length}
Focusing the electron beam in the center of the cavity significantly increases the size of the beam at the position of the slit. As a result, the expression for the pulse length of equation \eqref{tau1}, which was done for an infinitely small beam, is no longer valid. The actual temporal profile of the pulses behind the slit is now proportional to the convolution of the tophat distribution of the chopping aperture and the approximately tophat distribution of the electron beam at the position of the chopping aperture.

In a TEM, the rms divergence angle $\sigma_{x'}$ is defined by the diameter of the C2-aperture and its distance to the center of the cavity. In the special case that the cavity is placed exactly in between the C2-aperture and a circular chopping aperture; and both apertures have the same diameter, the resulting rms pulse length is given by
\begin{equation}\label{tau}
\sigma_t=\frac{\sqrt{2}\gamma_0 m \sigma_{x'}}{q B_0\sin(\Lambda/2)}.
\end{equation}
Equation \eqref{tau} shows that to maintain short pulses while focusing in the cavity, is it important to select the divergence angle as small as possible. Therefore, is it not only important to choose a small chopping aperture, but also a small C2-aperture, preferably of the same diameter.

\subsection{Energy spread}
The expression for the final energy spread  $\sigma_U^f$ that describes figure \ref{fig:emesgrowth} is obtained by evaluating equation \eqref{esfinal} and is even more complicated than the general expression for the final transverse emittance $\epsilon_x^f$. To gain insight in the different parameters contributing to a growth in energy spread, we consider three different situations in which we substitute three different sets of assumptions in equation \eqref{esfinal}.

\begin{itemize}
\item{
We start by assuming an ideal, focused beam, i.e. substituting $\epsilon_{n,x}^i=\sigma_U^i=\sigma_t=0$ in equation \eqref{esfinal}. We find
\begin{equation}
\sigma_U^f=\gamma_0 m v_0 \sigma_{x'} \frac{\omega_c}{\omega}\left| \frac{z_0}{L_{}} \Lambda \cos{\phi_0} - \left(\frac{z_0}{L_{}}-1\right)\Lambda \cos{(\Lambda+\phi_0)}+\sin{\phi_0}-\sin{(\Lambda+\phi_0)}  \right|.
\label{thetacontribution}
\end{equation}
This contribution is plotted in figure \ref{fig:emesgrowth}b as the white dashed curve. It has the same functional dependence on $\phi_0$ and $z_0$ as the increase in transverse emittance in equation \eqref{emgr}, and can be fully eliminated using the same conjugate blanking scheme of equation \eqref{conjugate}.
}
\item{
However, the energy spread also increases with the initial pulse duration $\sigma_t\neq0$. By substituting $\sigma_{x'}=0$, $\Lambda=\pi$ and $z_0=L/2$ into equation \eqref{esfinal}, instead of $\sigma_t=0$, we find
\begin{equation}\label{taucontribution}
\sigma_U^f=\gamma_0 m v_0^2 \pi  \frac{\omega_c^2}{ \omega} \sigma_t\sqrt{\cos\left(2 \phi_0^2\right)}.
\end{equation}
This contribution describes the black dashed lines in figure \ref{fig:emesgrowth}b and explains the minima in $\sigma_U^f$ at $\phi_0=\pm \frac{\pi}{4}$. At the intersection points of the black dashed lines and the white dashed curve, both the increase in transverse emittance and energy spread are eliminated simultaneously. In principle, we can exploit these 'sweet spots' by placing the slit off-axis such that phase $\phi_0=\pm \frac{\pi}{4}$ is chopped out of the beam, see also figure \ref{fig:trajectories}b, followed by choosing the correct position of the crossover. In practice, the deflection coils in a TEM could be used to redirect the beam back to the optical axis.
}
\item{
However, for easy switching between pulsed mode and continuous mode; and to obtain the shortest pulses, we choose to place the slit on-axis. For a beam with zero initial energy spread $\sigma_U^i=0$ and the chopping slit placed on-axis, hence $\phi_0=\frac{1}{2}(\pi-\Lambda)$, the final energy spread is given by
\begin{equation}
\boxed{
\sigma_U^f=\gamma_0^3 m v_0^2 \frac{\omega_c}{\omega}\sqrt{\left[4\sigma_{x'}^2\left(\frac{z_0}{L_{}}-\frac{1}{2}\right)^2\Lambda^2+\frac{k^2\left(\epsilon_x^i\right)^2}{\beta_0^2 \sigma_{x'}^2}\right]\sin^2{\left(\frac{\Lambda}{2}\right)}+ \frac{\omega_c^2\sigma_t^2}{\gamma_0^2}(\Lambda-\sin{\Lambda})^2}.
}
\label{esgr}
\end{equation}
Equation \eqref{esgr} shows three terms:
 \begin{itemize}
 \item{The first term is proportional to the angular divergence $\sigma_{x'}$ and can be eliminated by focusing in the center of the cavity, i.e. $z_0=\frac{L_{}}{2}$.}
      \item{The second term scales with $k \frac{\epsilon_x^i}{\sigma_{x'}}$ and describes the sampling of small, off-axis electric fields due to the finite size of the beam in the crossover $z=z_0$. It can be minimized by decreasing the transverse emittance of the incident beam, for instance by using a smaller C2-aperture at the expense of average current.}
           \item{The third term can be reduced by decreasing the cavity length parameter $\Lambda$. However, according to equation \eqref{tau1} this goes at the expense of the pulse length. When we decrease the cavity length parameter from $\Lambda=\pi$ to $\Lambda=\Lambda'$ while we simultaneously increase the magnetic field by a factor
\begin{equation}\frac{\omega_c'}{\omega_c}=\frac{\sin{\left(\pi/2\right)}}{\sin{\left(\Lambda'/2\right)}},
\end{equation}
the growth in energy spread can be decreased by a factor
\begin{equation}
\frac{\Delta \sigma_U'(\omega_c',\Lambda')}{\Delta \sigma_U(\omega_c,\pi)} =\frac{\Lambda'-\sin{\Lambda'}}{\pi \sin^2{\left(\frac{\Lambda'}{2}\right)}}<1, \mathrm{~ for ~ \Lambda'<\pi}
\end{equation}
while the short pulse length is fully maintained. Of course, the increased magnetic field in the cavity requires a higher input power. A second way to decrease the third contribution to $\sigma_U^f$ is to decrease the pulse length $\sigma_t$ by decreasing the divergence $\sigma_{x'}$ of the incident beam, see equation \eqref{tau}.}
\end{itemize}

Figure \ref{fig:cavitylengthdivergence}a shows the final energy spread $\sigma_U^f$ of equation \eqref{esgr} for a beam with no initial energy spread focused in the center of the cavity ($z_0=L/2$) as a function of divergence angle $\sigma_{x'}$ for varying cavity length parameter $\Lambda$ and initial emittance $\epsilon_x^i$. Here we have increased the magnetic field accordingly to keep the rms pulse length $\sigma_t = 100$ fs. This is done by solving equation \eqref{tau} for $B_0$ and substituting that in equation \eqref{esgr}. The magnetic field $B_0$ required to create $\sigma_t=100$ fs pulses for decreasing cavity length parameter $\Lambda$ is plotted in figure \ref{fig:cavitylengthdivergence}b  for varying rms divergence angle $\sigma_{x'}$.
\begin{figure}[htp]
    \centering
        \includegraphics[width=\textwidth]{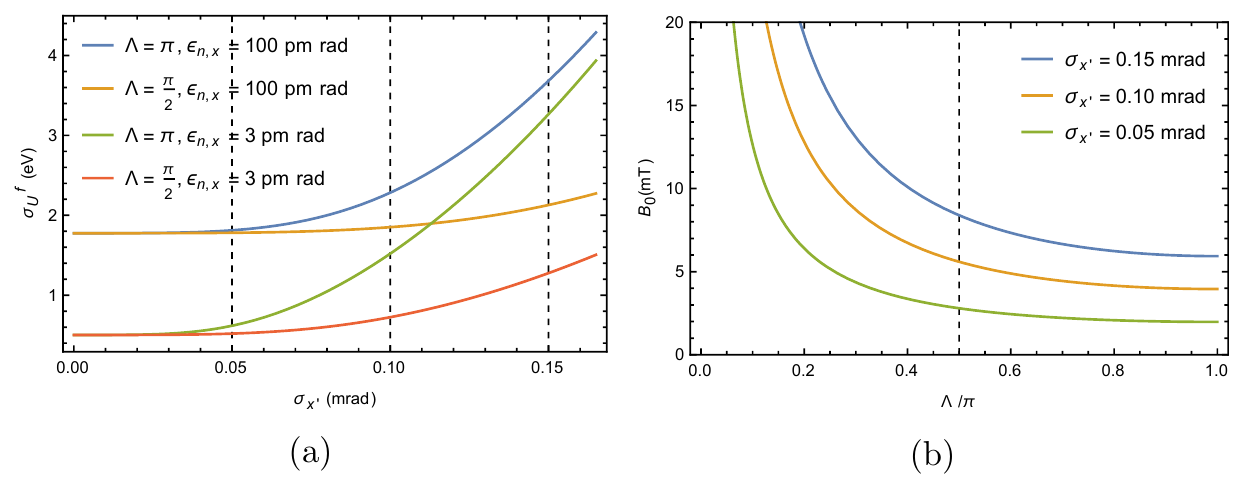}
        \label{fig:cavitylengthdivergence}
    \caption{Final energy spread $\sigma_U^f$ of equation \eqref{esgr} for a beam focused in the center of the cavity ($z_0=L/2$) as function of divergence angle $\sigma_{x'}$, for varying values of cavity length and initial emittance (a). To keep the pulse length fixed at $\sigma_t=100$ fs, the magnetic field amplitude $B_0$ in the cavity is altered accordingly (b).}
    \label{fig:cavitylengthdivergence}
\end{figure}

Figure \ref{fig:cavitylengthdivergence}a demonstrates two methods that can be used to minimize the final energy spread of the chopped pulses. First, the final energy spread can be reduced significantly by decreasing the cavity length while simultaneously increasing the magnetic field $B_0$ to maintain short pulses. This is an effective strategy down to a cavity length parameter of $\Lambda \approx \pi/2$, below which decreasing the cavity length further becomes very expensive in terms of $B_0$, see figure \ref{fig:cavitylengthdivergence}b. Secondly, decreasing the divergence angle $\sigma_{x'}$ can reduce the remaining growth in energy spread even further. This is effective down to the point where the transverse emittance of the incident beam $\epsilon_{n,x}^i$ limits the minimum achievable final energy spread.
}
\end{itemize}
\section{Particle tracking simulations}\label{sec:simulations}

In the previous sections, we have used an analytical approach to show that an ideal cylindrical cavity in TM$_{110}$ mode can be used to chop a relativistic electron beam into ultrashort electron pulses while maintaining the quality of the original beam. At this point, we want to investigate the limits of this technique in a scenario as realistic as possible. Therefore, we first implement the actual cavity geometry used in the TU/e UTEM in \textsc{cst} Microwave Studio \cite{Studio2006} and numerically calculate the realistic $B_y$-field and the $E_x$-field along the cavity axis. The diameters of both the entrance and the exit aperture are 3 mm.

\begin{figure}[htp]
\centering
        \includegraphics[width=\textwidth]{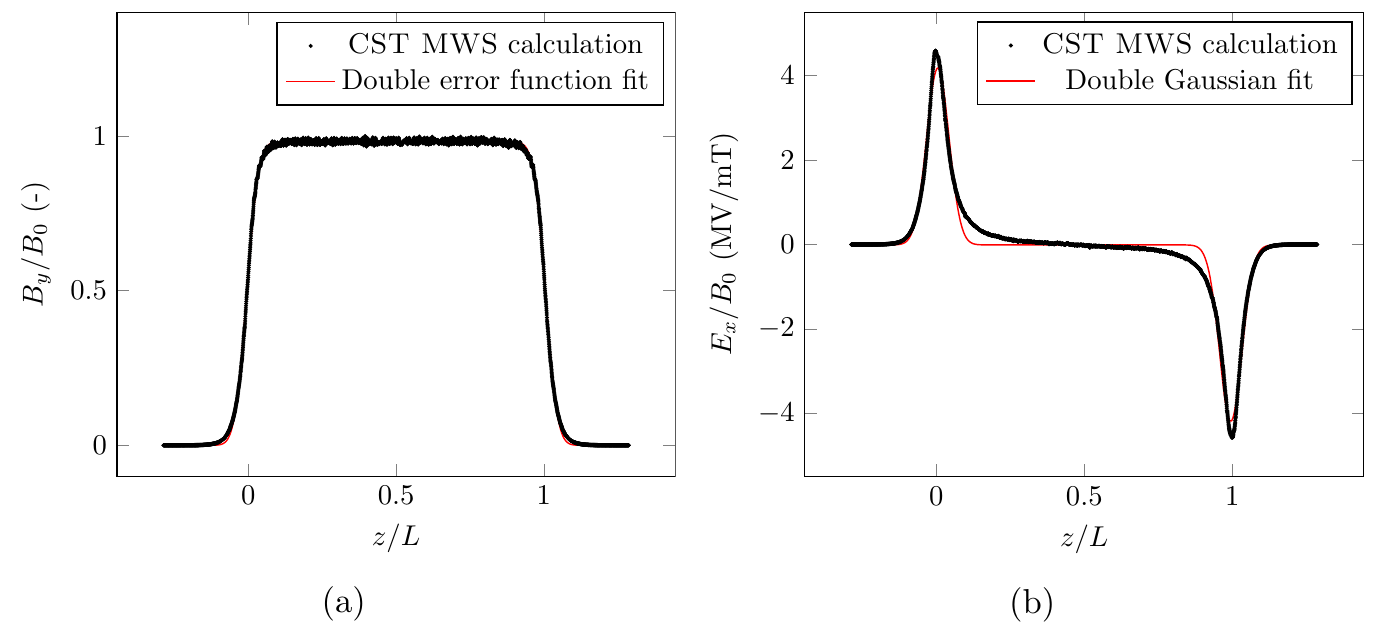}
    \caption{Scatter: \textsc{cst} Microwave Studio simulations of the $B_y(r=0,z)$-field (a) and $E_x(r=0,z)$-field (b) for a realistic TM$_{110}$ cavity geometry. Solid line: double error-function fit (a) and double Gaussian fit (b).}\label{fig:CST}
\end{figure}

Figure \ref{fig:CST} shows that the actual cavity geometry results in fringe fields near the entrance and exit apertures. Figure \ref{fig:CST}a shows the on-axis $B_y$-field, which is fitted with a double error-function
\begin{equation}
B_y(r=0,z)=\frac{B_{y,0}}{\mathrm{erf}(\frac{L_{}}{2s})}\left(\frac{1}{2}\mathrm{erf}\left(\frac{z}{s}\right)- \frac{1}{2}\mathrm{erf}\left(\frac{z-L_{}}{s}\right)\right).
\end{equation}
Here $B_{y,0}$ is the maximum field strength at the center of the cavity and $s$ is a fit-parameter that describes how the $B_y$-field falls off near the apertures of the cavity. Figure \ref{fig:CST}b shows the on-axis $E_x$-field, which is fitted with two Gaussians of opposite sign. When we substitute these fits into a fifth order power expansion of the solution of a cylindrical cavity, see equation \eqref{fifth} in appendix A, we can reconstruct all the other $(r,\varphi,z)$ components of the EM field close to the cavity axis. To investigate whether the fringe fields near the cavity apertures affect the beam quality, we implement the obtained field expansions of equation \eqref{fifth} in the \textsc{gpt}-code \cite{VanderGeer2001} for realistic particle tracking simulations. These simulations also allow us to chop a continuous electron beam using the combination of a cavity and a slit, rather than ab initio assuming a Gaussian temporal distribution. Furthermore, we can simulate a more realistic initial beam with uniform spatial and angular distributions, rather than Gaussian distributions.

In the simulations, we apply the practical lessons we learned from the Courant-Snyder model in the particle tracking simulations:
\begin{itemize}
\item{We select RF phase $\phi_0=\frac{1}{2}(\pi-\Lambda)$ by placing the slit on-axis to obtain the shortest pulses.}
\item{We focus in the center of the cavity $z_0=L/2$ to reduce the growth in transverse emittance and energy spread.}
\item{We choose the cavity length parameter $\Lambda=\pi/2$ to reduce the remaining growth in energy spread even further while still being able to make short pulses with a realistic cavity field amplitude.}
\end{itemize}
For the electron source we choose a typical 200 kV Schottky field-emission gun with a practical reduced brightness $B_r=10^8$ A/m$^2$ sr V and rms energy spread $\sigma_U^i=0.5$ eV. A DC current of $I=10$ nA then results in an initial normalized transverse emittance $\epsilon_{n,x}^i = 3$ pm rad (rms). To test the validity of our theory for applications with higher currents, we also add a simulation series for $\epsilon_{n,x}^i = 100$ pm rad (rms). Furthermore, we vary the magnetic field amplitude from $B_0=1$ mT to $B_0=10$ mT in steps of 1 mT and we vary the diameter of the C2- and chopping apertures between $d=30$ $\mu$m and $d=10$ $\mu$m. For these parameter settings, we measure the final rms pulse length $\sigma_t$, the rms normalized transverse emittance $\epsilon_{n,x}^f$ and rms energy spread $\sigma_U^f$.

\begin{figure}[htp]
\centering
\includegraphics[width=\linewidth]{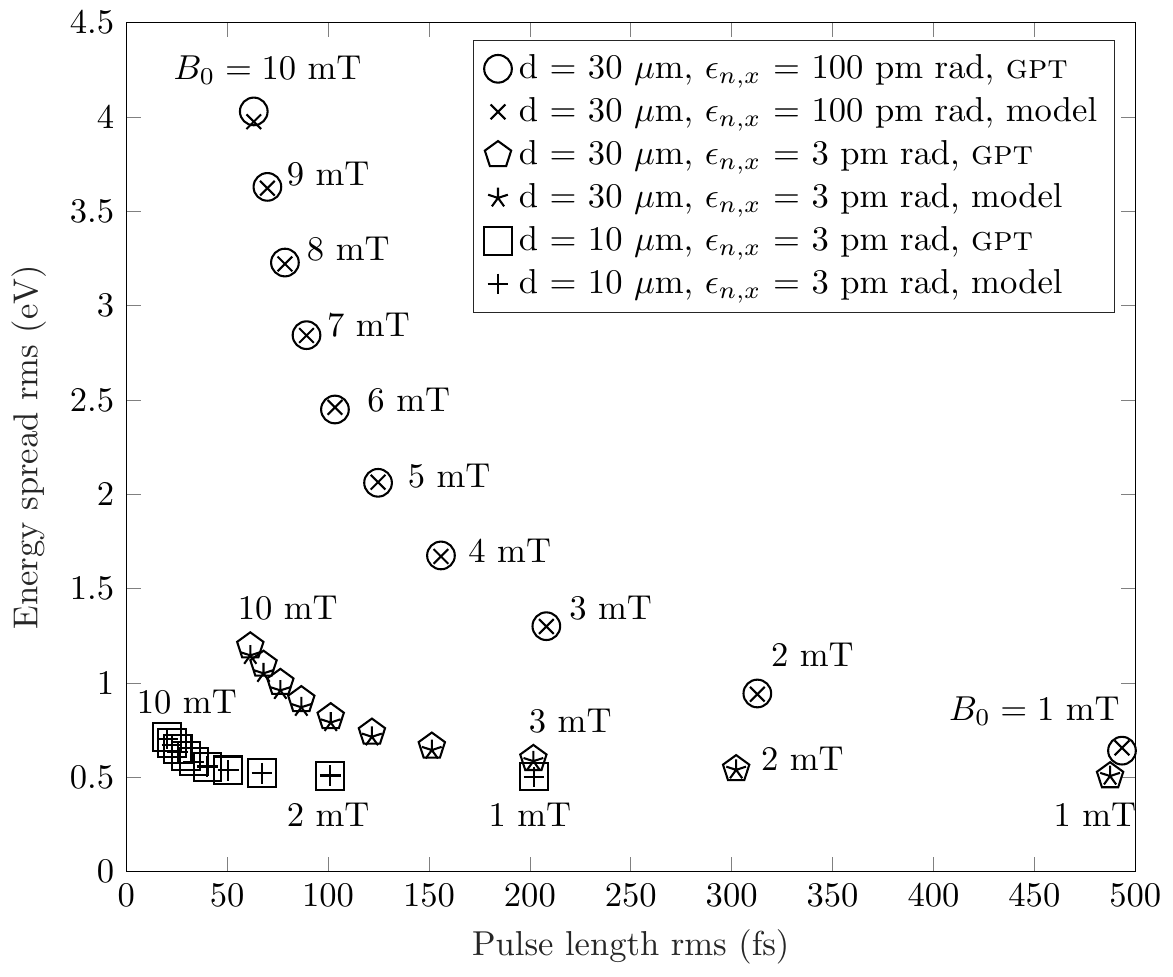}
\caption{\textsc{gpt}-simulations ($\Circle,\pentagon,\Square$) and results of the Courant-Snyder model ($\times,\star,+$) of the final rms energy spread $\sigma_U^f$ and final rms pulse length $\sigma_t$ of a spatially and temporally uniform electron beam (initial normalized rms emittance $\epsilon_{n,x}$ and initial rms energy spread $\sigma_U^i=0.5$ eV) focused in the center of a TM$_{110}$ cavity and chopped by an on-axis aperture of diameter $d$, for various values of the cavity field amplitude $B_0=1,2,...,10$ mT.}
\label{fig:GPTsimulation}
\end{figure}

Figure \ref{fig:GPTsimulation} shows both the results of these particle tracking simulations and the results of the Courant-Snyder model. The latter have been obtained by substituting the simulated rms values for $\sigma_{x'}$ and $\sigma_t$ in the non-simplified version of equation \eqref{esgr}. First of all we find excellent agreement between the theoretical model and the particle tracking simulations. This gives good confidence that the derived expressions for the optical transfer matrix correctly describe the actual particle trajectories, and therefore justifies the perturbative approach of describing a TM$_{110}$ cavity as a linear optical element for electrons based on the assumptions in section \ref{sec:framework}. Furthermore it shows that the fringe fields due to a non-idealized cavity geometry do not significantly affect the beam dynamics.
Secondly, figure \ref{fig:GPTsimulation} shows that there is a trade off between the final energy spread of the electron beam and the final pulse length. The amplitude of the magnetic field in the cavity can be used to shape the final time-energy phase space distribution, depending on the application. However above all, figure \ref{fig:GPTsimulation} demonstrates the enormous potential for RF cavity based ultrafast electron microscopy. Especially when 10 $\mu$m apertures are used in combination with a low emittance input beam of 3 pm rad, extremely short pulses can be generated with hardly any increase in energy spread. Pulses of 200 fs (rms) with $\sigma_U^f=0.5$ are expected even for a magnetic field amplitude of only $B_0=1$ mT, which has already been realized at TU/e \cite{Verhoeven2017}. Ultimately, pulse lengths down to 20 fs (rms), combined with rms energy spread of $\sigma_U^f=0.70$ eV are expected for $B_0=10$ mT. Finally, we didn't observe any increase in normalized transverse emittance in any of the simulations in figure \ref{fig:GPTsimulation}.

As for any pulsed beam, a short pulse length results in a low average current of the pulsed beam. Although the 3 GHz repetition rate of an RF cavity based UEM is several orders of magnitude higher than in conventional UEMs based on photo-emission, the low charge per pulse: $0.006$ e/pulse at $I=10$ nA for $\tau = 100$ fs, limits the average current. However, because the peak brightness of the original beam is conserved after chopping with an RF cavity, any future improvements on the continuous source will directly improve a cavity-based pulsed source as well. For example, the recent developments of new LaB$_6$-emitters which promise a brightness up to $B_r=10^{10}$ A/m$^2$sr eV \cite{Zhang2015} are worth mentioning. Provided that they can be operated at sufficient current, the combination of such sources with an RF cavity could result in ultrafast electron microscopy with unprecedented spatial and temporal resolution at the average current of present-day continuous electron microscopes.

\section{Conclusions}\label{sec:conclusions}
We have developed a theoretical description of resonant radiofrequency deflecting cavities in TM$_{110}$ mode as dynamic optical elements for ultrafast electron microscopy. We have derived the optical transfer matrix of an ideal pillbox cavity and have calculated the 6D phase space propagation of a Gaussian electron distribution using a Courant-Snyder formalism. We have derived closed, analytic expressions for the increase in transverse emittance and energy spread that have resulted in practical insight that can be applied directly in an experiment. We have shown that the beam quality of the incident electron beam can by maintained by proper settings of the RF phase and the position of the crossover inside the cavity. In particular, we have explained the concept of conjugate blanking for fully eliminating increase in transverse emittance. The growth in energy spread can be minimized by decreasing the cavity length and the divergence angle. The correctness of our model and the potential of RF cavities for UEM are confirmed by charged particle tracking simulations using a realistic cavity geometry, that take into account fringe fields at the cavity entrance and exit apertures. In conclusion, RF cavities in TM$_{110}$ mode allow high-repetition rate, ultrafast electron microscopy with 100 fs temporal resolution combined with the atomic resolution of a high-end TEM.

\section*{Acknowledgement}
This work is part of an Industrial Partnership Programme of the Netherlands Organisation for Scientific Research (NWO).

\section*{References}
\bibliographystyle{unsrt}
\bibliography{bibliography}

\begin{thebibliography}{10}

\bibitem{Lobastov2005}
V.A. Lobastov, R.~Srinivasan, and A.H. Zewail.
\newblock Four-dimensional ultrafast electron microscopy.
\newblock {\em Proceedings of the National Academy of Sciences of the United
  States of America}, 102(20):7069--7073, 2005.

\bibitem{Zewail2010}
A.H. Zewail.
\newblock Four-dimensional electron microscopy.
\newblock {\em Science}, 328(5975):187--193, 2010.

\bibitem{Flannigan2012}
D.J. Flannigan and A.H. Zewail.
\newblock 4{D} electron microscopy: principles and applications.
\newblock {\em Acc.~Chem.~Res.}, 45:1828--1839, 2012.

\bibitem{Sciaini2011}
G.~Sciaini and R.J.D. Miller.
\newblock Femtosecond electron diffraction: heralding the era of atomically
  resolved dynamics.
\newblock {\em Reports on Progress in Physics}, 74(9):096101, 2011.

\bibitem{Carbone2009}
F.~Carbone, B.~Barwick, O.~Kwon, H.~Soon Park, J.S. Baskin, and A.H. Zewail.
\newblock {EELS} femtosecond resolved in 4{D} ultrafast microscopy.
\newblock {\em Chem.~Phys.~Lett.}, 468:4, 2008.

\bibitem{VanderVeen2015}
R.M. van~der Veen, T.J. Penfold, and A.H. Zewail.
\newblock Ultrafast core-loss spectroscopy in four-dimensional electron
  microscopy.
\newblock {\em Struct.~Dyn.}, 2:024302, 2015.

\bibitem{Feist2017}
A.~Feist, N.~Bach, N.~R. da~Silva, T.~Danz, M.~M{\"o}ller, K.E. Priebe,
  T.~Domr{\"o}se, G.J. Gatzmann, S.~Rost, J.~Schauss, et~al.
\newblock Ultrafast transmission electron microscopy using a laser-driven field
  emitter: femtosecond resolution with a high coherence electron beam.
\newblock {\em Ultramicroscopy}, 2016.

\bibitem{Ehberger2015}
D.~Ehberger, J.~Hammer, M.~Eisele, M.~Kr\"uger, J.~Noe, A.~H\"ogele, and
  P.~Hommelhoff.
\newblock Highly coherent electron beam from a laser-triggered tungsten needle
  tip.
\newblock {\em Phys. Rev. Lett.}, 114:227601, Jun 2015.

\bibitem{Feist2015}
A.~Feist, K.E. Echternkamp, J.~Schauss, S.V. Yalunin, S.~Schafer, and
  C.~Ropers.
\newblock Quantum coherent optical phase modulation in an ultrafast
  transmission electron microscope.
\newblock {\em Nature}, 521:7551, 2015.

\bibitem{Echternkamp2016}
K.E. Echternkamp, A.~Feist, S.~Sch{\"a}fer, and C.~Ropers.
\newblock Ramsey-type phase control of free-electron beams.
\newblock {\em Nature Physics}, 12(11):1000--1004, 2016.

\bibitem{Fujioka1983}
H.~Fujioka and K.~Ura.
\newblock Electron beam blanking systems.
\newblock {\em Scanning}, 5(1):3--13, 1983.

\bibitem{Thong1991}
J.T.L. Thong.
\newblock Picosecond electron pulse generation via beam deflection-chopping in
  the sem.
\newblock {\em Measurement Science and Technology}, 2(3):207, 1991.

\bibitem{Winkler1990}
D.~Winkler, R.~Schmitt, M.~Brunner, and B.~Lischke.
\newblock Flexible picosecond probing of integrated circuits with chopped
  electron beams.
\newblock {\em IBM journal of research and development}, 34(2.3):189--203,
  1990.

\bibitem{Fehr1990}
J.~Fehr, W.~Reiners, L.J. Balk, E.~Kubalek, D.~K{\"o}ther, and I.~Wolff.
\newblock A 100-femtosecond electron beam blanking system.
\newblock {\em Microelectronic Engineering}, 12(1-4):221--226, 1990.

\bibitem{Oldfield2001}
L.~Oldfield.
\newblock A rotationally symmetric electron beam chopper for picosecond pulses.
\newblock {\em J.~Phys.~E: Sci.~Instrum.}, 9:6, 1976.

\bibitem{Hosokawa1978}
T.~Hosokawa, H.~Fujioka, and K.~Ura.
\newblock Generation and measurement of subpicosecond electron beam pulses.
\newblock {\em Rev.~Sci.~Instrum.}, 49:624, 1978.

\bibitem{Weppelman2018}
I.G.C. Weppelman, R.J. Moerland, J.P. Hoogenboom, and P.~Kruit.
\newblock Concept and design of a beam blanker with integrated photoconductive
  switch for ultrafast electron microscopy.
\newblock {\em Ultramicroscopy}, 184(Part B):8 -- 17, 2018.

\bibitem{Kiewiet2002}
F.B. Kiewiet, A.H. Kemper, O.J. Luiten, G.J.H. Brussaard, and M.J. van~der
  Wiel.
\newblock Femtosecond synchronization of a 3ghz rf oscillator to a mode-locked
  ti: sapphire laser.
\newblock {\em Nuclear Instruments and Methods in Physics Research Section A:
  Accelerators, Spectrometers, Detectors and Associated Equipment},
  484(1):619--624, 2002.

\bibitem{Gliserin2013}
A.~Gliserin, M.~Walbran, and P.~Baum.
\newblock Passive optical enhancement of laser-microwave synchronization.
\newblock {\em Applied Physics Letters}, 103(3):031113, 2013.

\bibitem{Brussaard2013}
G.J.H. Brussaard, A.C. Lassise, P.L.E.M. Pasmans, P.H.A. Mutsaers, M.J. van~der
  Wiel, and O.J. Luiten.
\newblock Direct measurement of synchronization between femtosecond laser
  pulses and a 3 {GHz} radio frequency electric field inside a resonant cavity.
\newblock {\em Appl.~Phys.~Lett.}, 103:141105, 2013.

\bibitem{Walbran2015}
M.~Walbran, A.~Gliserin, K.~Jung, J.~Kim, and P.~Baum.
\newblock 5-femtosecond laser-electron synchronization for pump-probe
  crystallography and diffraction.
\newblock {\em Physical Review Applied}, 4(4):044013, 2015.

\bibitem{VanOudheusden2010}
T.~Van~Oudheusden, P.L.E.M. Pasmans, S.B. Van Der~Geer, M.J. De~Loos, M.J. Van
  Der~Wiel, and O.J. Luiten.
\newblock Compression of subrelativistic space-charge-dominated electron
  bunches for single-shot femtosecond electron diffraction.
\newblock {\em Physical review letters}, 105(26):264801, 2010.

\bibitem{Lassise2012}
A.C. Lassise, P.H.A. Mutsaers, and O.J. Luiten.
\newblock Compact, low power radio frequency cavity for femtosecond electron
  microscopy.
\newblock {\em Review of Scientific Instruments}, 83(4):043705, 2012.

\bibitem{Verhoeven2016}
W.~Verhoeven, J.F.M. van Rens, M.A.W. van Ninhuijs, W.F. Toonen, E.R. Kieft,
  P.H.A. Mutsaers, and O.J. Luiten.
\newblock Time-of-flight electron energy loss spectroscopy using {TM}$_{110}$
  deflection cavities.
\newblock {\em Structural Dynamics}, 3:054303, 2016.

\bibitem{Franssen2017}
J.G.H. Franssen, T.L.I. Frankort, E.J.D. Vredenbregt, and O.J. Luiten.
\newblock Pulse length of ultracold electron bunches extracted from a laser
  cooled gas.
\newblock {\em Structural Dynamics}, 4(4):044010, 2017.

\bibitem{Kealhofer2016}
C.~Kealhofer, W.~Schneider, D.~Ehberger, A.~Ryabov, F.~Krausz, and P.~Baum.
\newblock All-optical control and metrology of electron pulses.
\newblock {\em Science}, 352(6284):429--433, 2016.

\bibitem{Lassise2012PhDthesis}
A.C. Lassise.
\newblock {\em Miniaturized RF technology for femtosecond electron microscopy}.
\newblock PhD thesis, Ph. D. thesis, Eindhoven University of Technology, 2012.

\bibitem{Verhoeven2017}
W.~Verhoeven, J.F.M. van Rens, E.R. Kieft, P.H.A. Mutsaers, and O.J. Luiten.
\newblock High quality ultrafast transmission electron microscopy using
  resonant microwave cavities.
\newblock {\em arXiv preprint arXiv:1709.02205}, 2017.

\bibitem{Qiu}
J.~Qiu, G.~Ha, C.~Jing, S.V. Baryshev, B.W. Reed, J.W. Lau, and Y.~Zhu.
\newblock Ghz laser-free time-resolved transmission electron microscopy: A
  stroboscopic high-duty-cycle method.
\newblock {\em Ultramicroscopy}, 161:130--136, 2016.

\bibitem{Reiser}
M.~Reiser.
\newblock {\em Theory and design of charged particle beams}.
\newblock John Wiley \& Sons, second edition, 2008.

\bibitem{Bronsgeest2008}
M.S. Bronsgeest, J.E. Barth, L.W. Swanson, and P.~Kruit.
\newblock Probe current, probe size, and the practical brightness for probe
  forming systems.
\newblock {\em Journal of Vacuum Science \& Technology B: Microelectronics and
  Nanometer Structures Processing, Measurement, and Phenomena}, 26(3):949--955,
  2008.

\bibitem{Courant1958}
E.D. Courant and H.S. Snyder.
\newblock Theory of the alternating-gradient synchrotron.
\newblock {\em Annals of physics}, 281(1-2):360--408, 2000.

\bibitem{Studio2006}
Microwave Studio.
\newblock Cst-computer simulation technology.
\newblock {\em Bad Nuheimer Str}, 19:64289, 2008.

\bibitem{VanderGeer2001}
S.B. Van~der Geer and M.J. de~Loos.
\newblock The general particle tracer code: design, implementation and
  application.
\newblock {\em PhD Thesis}, 2001.

\bibitem{Zhang2015}
H.~Zhang, J.~Tang, J.~Yuan, Y.~Yamauchi, T.T. Suzuki, N.~Shinya, K.~Nakajima,
  and L.~Qin.
\newblock An ultrabright and monochromatic electron point source made of a
  {L}a{B}6 nanowire.
\newblock {\em Nature nanotechnology}, 11(3):273--279, 2016.

\end{thebibliography}

\section*{Appendix A: Fifth order power expansion of EM fields in a cylindrical cavity in TM$_{110}$ mode}\label{sec:xavier}

The fifth order power expansion of the $(r,\varphi,z)$ components of the EM field in a cylindrical cavity TM$_{110}$ mode are given by:
\begin{equation}\label{fifth}
\begin{aligned}
E_r =&\frac{1}{192}  \left\{(192-24k^2 r^2+k^4 r^4 ) E_x (0,z)+(-72r^2+6k^2 r^4 )  \frac{\partial^2 E_x (0,z)}{\partial^2 z}+5r^4  \frac{\partial^4 E_x (0,z)}{\partial^4 z}\right.\\
&\left.+(-48r^2+4k^2 r^4 )kc \frac{\partial B_y (0,z)}{\partial z}+4r^4 kc \frac{\partial^3 B_y (0,z)}{\partial^3 z}\right\}\cos⁡{\varphi} \cos⁡(\omega t+\phi_0)\\
E_\varphi =&\frac{1}{192}  \left\{(-192+72k^2 r^2-5k^4 r^4 ) E_x (0,z)+(24r^2-6k^2 r^4 )  \frac{\partial^2 E_x (0,z)}{\partial^2 z}-r^4  \frac{\partial^4 E_x (0,z)}{\partial^4 z}\right.\\
&\left.+(-48r^2+4k^2 r^4 )kc \frac{\partial B_y (0,z)}{\partial z}+4r^4 kc \frac{\partial^3 B_y (0,z)}{\partial^3 z}\right\}\sin⁡{\varphi} \cos⁡(\omega t+\phi_0)\\
E_z =&\frac{1}{192} \cos⁡\varphi \cos⁡(\omega t+\phi_0) \left\{(192r-24k^2 r^3+k^4 r^5 )\left(\frac{\partial E_x (0,z)}{\partial z}+kcB_y (0,z)\right)\right.\\
&\left.+(-24r^3+2k^2 r^5 )\left(\frac{\partial^3 E_x (0,z)}{\partial^3 z}+kc \frac{\partial^2 B_y (0,z)}{\partial^2 z}\right)+r^5 \left(\frac{\partial^5 E_x (0,z)}{\partial^5 z}+kc \frac{\partial^4 B_y (0,z)}{\partial^4 z}\right)\right\}\\
B_r =&\frac{1}{192} \left\{(192-24k^2 r^2+k^4 r^4 ) B_y (0,z)+(-72r^2+6k^2 r^4 )  \frac{\partial^2 B_y (0,z)}{\partial^2 z}+5r^4  \frac{\partial^4 B_y (0,z)}{\partial^4 z}\right.\\
&\left.+(48r^2-4k^2 r^4 )  \frac{k}{c}  \frac{\partial E_x (0,z)}{\partial z}-4r^4  \frac{k}{c}  \frac{\partial^3 E_x (0,z)}{\partial^3 z}\right\}\sin⁡\varphi \sin⁡(\omega t+\phi_0) \\
B_\varphi =&\frac{1}{192} \left\{(192-72k^2 r^2+5k^4 r^4 ) B_y (0,z)+(-24r^2+6k^2 r^4 )  \frac{\partial^2 B_y (0,z)}{\partial^2 z}+r^4  \frac{\partial^4 B_y (0,z)}{\partial^4 z}\right.\\
&\left.+(-48r^2+4k^2 r^4 )  \frac{k}{c}  \frac{\partial E_x (0,z)}{\partial z}+4r^4  \frac{k}{c}  \frac{\partial^3 E_x (0,z)}{\partial^3 z}\right\} \cos⁡\varphi \sin⁡(\omega t+\phi_0)\\
B_z =&\frac{1}{192}  \sin⁡\varphi \sin⁡(\omega t+\phi_0) \left\{(192r-24k^2 r^3+k^4 r^5 )\left(\frac{\partial B_y (0,z)}{\partial z}-\frac{k}{c} E_x (0,z)\right)\right.\\
&\left.+(-24r^3+2k^2 r^5 )\left(\frac{\partial^3 B_y (0,z)}{\partial^3 z}-\frac{k}{c} \frac{\partial^2 E_x (0,z)}{\partial^2 z}\right)+r^5 \left(\frac{\partial^5 B_y (0,z)}{\partial^5 z}-\frac{k}{c}  \frac{\partial^4 E_x (0,z)}{\partial^4 z}\right)\right\},
\end{aligned}
\end{equation}
in which $E_x(0,z)$ and $B_y(0,z)$ are the on-axis electric and magnetic field amplitudes. Describing the cavity using these field expansions rather than a 6D field map results into 10-20 times faster particle tracking simulations.

\end{document}